\documentclass{pasj00}
\draft

\begin{document}
\SetRunningHead{Y. Fujita}{Pre-Processing of Galaxies}
\Received{2003/10/03}
\Accepted{2003/11/01}

\title{Pre-Processing of Galaxies before Entering a Cluster}

\author{Yutaka \textsc{Fujita}}
\affil{National Astronomical
Observatory, \\and Department
of Astronomical Science, The Graduate University for Advanced Studies,\\
Osawa 2-21-1, Mitaka, Tokyo 181-8588}
\email{yfujita@th.nao.ac.jp}

\KeyWords{galaxies: clusters: general---galaxies: evolution---galaxies:
high-redshift---galaxies: interactions}

\maketitle

\begin{abstract}
 We consider several mechanisms that possibly affect the evolution of
 disk galaxies in clusters using analytical models based on a
 hierarchical clustering scenario.  We especially focus on the evolution
 of disk galaxies in subclusters located around a main cluster. We show
 that ram-pressure stripping cannot be always ignored in subclusters,
 although their masses are much smaller than that of the main
 cluster. The star-formation rate of a galaxy may gradually decrease by
 the stripping of warm gas (`strangulation') in a main cluster. However,
 we find that ram-pressure stripping could start before the
 strangulation is completed, if a field galaxy directly falls into the
 main cluster.  Since this conflicts with some recent observations, many
 galaxies might have been affected by some environmental effects when
 they were in subclusters before they fell into the main cluster
 (`pre-processing'). We show that strangulation and evaporation of the
 cold gas by the surrounding hot ICM in subclusters are the possible
 candidates.  We also show that the observed morphological
 transformation of disk galaxies at $z\lesssim 1$ is not chiefly due to
 galaxy mergers.
\end{abstract}

\section{Introduction}

Clusters of galaxies in the redshift range of $z\gtrsim 0.2$ often
exhibit an overabundance, relative to present-day clusters, of blue
galaxies (Butcher, Oemler 1978, 1984). This star-formation activity is
often called the Butcher--Oemler effect, and subsequent studies have
confirmed this trend \citep{cou87,rak95,lub96,mar00,ell01,got03a}. At
the same time, various studies have reported a morphological evolution
in cluster galaxies; they have shown that the fractions of S0 galaxies
in clusters decrease toward a higher redshift (Dressler et al. 1997;
Couch et al. 1994, 1998; Fasano et al. 2000). For many clusters, it has
been found that their galaxy population gradually changes from a red,
evolved, early type population in the inner part of the clusters to a
progressively blue, later type population in the extensive outer
envelope of the clusters
\citep{abr96,bal97,rak97,oem97,sma98,cou98,van98}. These are often
interpreted as being the result that star-forming field disk galaxies
infalling to a cluster are transformed to passively evolving cluster
members.

Various mechanisms have been proposed for the transformation. In
general, galaxies in the inner region of a cluster have lower
star-formation rates both at $z\sim 0$ \citep{lew02,gav02,gra03} and at
higher redshift (Couch, Sharples 1987; Balogh et al. 1997, 1999b;
Poggianti et al. 1999; Postman et al. 2001; Couch et al. 2001; Balogh et
al. 2002c). Ram-pressure stripping of cold gas in
the disk of a galaxy is a possible candidate for the suppression of star
formation in clusters because the removal of cold gas leads to the
exhaustion of an ingredient of stars
\citep{gun72,bal94,fuj99a,aba99,qui00,mor00,vol01,bic02}, although
ram-pressure may cause a mild increase of star formation of a galaxy
\citep{fuj99a} or a starburst with a short-time scale
\citep{bek03,gav03}. Ram-pressure stripping is likely to take place in
the central regions of clusters because the density of intracluster
medium (ICM) is high there. In fact, disk galaxies with a deficit of
cold gas or a morphological sign of gas removal are often observed near
cluster centers
(Cayatte et al. 1990; Vollmer et al. 2000; Bravo-Alfaro et al. 2000,
2001; Solanes et al. 2001; Bureau, Carignan 2002; Vollmer 2003; Sofue et
al. 2003; Yoshida et al. 2003).  Moreover, it is more effective in
richer clusters because the typical velocity of galaxies is
larger. Several authors actually indicated that the fractions of blue
galaxies are systematically lower for rich clusters than for poor
clusters at a given redshift \citep{mar01,got03a}, which is consistent
with a suppression of the star-formation activities by ram-pressure
stripping. On the other hand, \citet{bal02b} investigated clusters with
small masses at $z\sim 0.25$, and indicated that the striking similarity
between the spectral and morphological properties of galaxies in these
clusters and those of galaxies in more massive systems at similar
redshifts implies that the physical processes responsible for truncating
star formation in galaxies are not restricted to the rare, rich cluster
environment, but are viable in much more common environments. Thus, they
concluded that ram-pressure stripping cannot be solely responsible for
the low star-formation rates in these systems. Moreover, several authors
indicated that ram-pressure stripping cannot account for the suppression
of star formation observed $\sim$Mpc away from the center of a cluster
\citep{bal97,kod01c,lew02}, although we will show that it is not
obvious.

\citet{lar80} suggested that an infall from gaseous galactic halos might
be important for sustaining star formation in spiral galaxies and that
the gas in these halos might be stripped in the cluster environment,
leading to the formation of passively evolving S0 galaxies. Although the
gas in galactic halos, which we call `warm gas', has not been detected
in nearby spiral galaxies \citep{ben00}, models of galaxy formation
based on the hierarchical clustering scenario have often assumed such an
infall, and have been very successful in accounting for the properties
of galaxies at optical and infrared wavelengths
\citep{col94,bau96,kau99,nag01}. If only warm gas is stripped, star
formation may be allowed to continue by consuming the remaining cold
disk gas, but without the infall to replenish this supply, star
formation will die out on timescales of a few Gyr \citep{lar80}. This is
consistent with observations that indicate a slow decline of the
star-formation rates of galaxies \citep{bal99b,kod01a}. We call this
scenario `strangulation' from now on. In this paper, we also consider
the evaporation of cold gas in disk galaxies by energy transfer via
thermal conduction from the surrounding hot ICM as another mechanism for
the slow decline of the star-formation rates. As far as we know, this is
the first time that the redshift evolution of the evaporation effect has
been studied.

The above mechanisms, that is, ram-pressure stripping, strangulation,
and evaporation, mainly have an influence on the star-formation
activities of galaxies. Contrary to them, galaxy mergers are expected to
significantly affect the morphology of galaxies. Thus, mergers may have
played an important role in the observed morphological transformation in
clusters. Mergers between galaxies with comparable masses (major
mergers) could create elliptical galaxies \citep{too72}. However, major
mergers alone cannot account for the observed fractions of galaxies with
intermediate bulge-to-disk luminosity ratios such as S0 galaxies in
clusters \citep{oka01,dia01}. Okamoto and Nagashima (2003) indicated
that mergers between galaxies with significantly different masses (minor
mergers) may produce the galaxies with intermediate bulge-to-disk
luminosity ratios because these mergers do not disrupt galactic disks
completely.

In this paper, we consider possible mechanisms responsible for the
truncation of star formation and morphological transformation of disk
galaxies in clusters. In particular, we investigate the evolution of
disk galaxies in subclusters. Recent observations show that galaxies
have already been affected by some environmental effects in subclusters
\citep{kod01c,tre03,got03b}. We define subclusters as clusters located
around a larger cluster (main cluster) at relatively high redshift;
these subclusters are to be merged with the main cluster until
$z=0$. Thus, at high redshift, we can say that these subclusters are
progenitors of the main cluster at $z=0$. On the other hand, at low
redshift, they become subhalos included in the main cluster. Since
high-resolution $N$-body simulations showed that a subcluster that has
been included in the main cluster as a subhalo is not completely
destroyed \citep{moo99,oka99,ghi00,fuk01}, we do not explicitly
discriminate a subcluster from a subhalo. However, we can estimate the
time when subclusters are included by the main cluster (see section
\ref{sec:result}).

We consider the effect of cosmological evolution of clusters since the
inner structure of clusters at high redshift is not the same as that at
low redshifts, which makes differences of environmental effects on
cluster galaxies at between high and low redshift. We take an analytical
approach to study the environmental effects because at the moment it is
difficult to study their redshift evolutions by genuine numerical
simulations. For example, it is virtually impossible to study
ram-pressure stripping of galaxies in clusters by cosmological numerical
simulations including both dark matter and gas with a resolution that is
sufficient to treat the interaction between ICM and interstellar gas in
each galaxy correctly. Even if we introduce some assumptions about the
stripping of the interstellar gas and the distribution of the ICM, as is
done by Okamoto and Nagashima (2003), we need to find the average
cluster and galaxy evolutions at least in terms of their dark
halos. However, it is time-consuming to create many clusters by
performing high-resolution $N$-body simulations to follow the average
cluster and galaxy evolutions. Moreover, there is another merit for
analytical approaches. Even in the future, when full numerical
simulations are enabled, the simulations must include many physical
processes, and it would be difficult to divide them when the results are
analyzed. The division is easier for analytical studies, and the results
of analytical studies would be very useful to be compared with those of
numerical simulations.  We also emphasize that our model is different
from previous semi-analytic models. Our model takes account of spatial
correlations among initial density fluctuations, and can specifically
predict the evolution of subclusters around the main cluster. On the
other hand, previous models cannot discriminate between subclusters and
the main cluster.

We mainly consider massive disk galaxies since they can be observed in
detail even at high redshift. Thus, we do not consider the cumulative
effect of high-speed encounters between galaxies (`galaxy harassment')
because it influences disk galaxies with small masses \citep{moo96}. We
mostly study the redshift evolution of each environmental effect and
discuss the relative strength of the effect at high and low redshift. We
also compare results with observations qualitatively.

This paper is organized as follows. In section~\ref{sec:model}, we
summarize our models. In section~\ref{sec:result} we give the results of
our calculations, and compare them with observations in
section~\ref{sec:disc}. Summary and conclusions are given in
section~\ref{sec:sum}. Readers who are interested only in the results of
calculations may skip section~\ref{sec:model}.

\section{Models}
\label{sec:model}

\subsection{The Growth of Clusters}
\label{sec:growth}

In this study, the average mass of progenitors of a cluster is derived
from the extend Press--Schechter model (EPS) \citep{bow91,bon91,lac93}
and its further extention \citep{fuj02b}. The latter is a
Press--Schechter model including the effect of spatial correlations
among initial density fluctuations (SPS). We briefly explain the model
here; the details are shown in \citet{fuj02b}.

We can estimate the conditional probability, $P(r,M_1,M_2)$, of finding
a region of mass $M_1$ with $\delta_1\geq\delta_{\rm c1}$ at a distance
$r$ from the center of an isolated, finite-sized object of mass $M_2$,
provided that the object of mass $M_1$ is included in the object of mass
$M_2$ $(>M_1)$ with $\delta_2=\delta_{\rm c2}$ at $r=0$. Here, we define
$\delta_M$ as the smoothed linear density fluctuation of mass scale $M$,
and $\delta_1=\delta_{M_1}$ and $\delta_2=\delta_{M_2}$.  Moreover, we
define $\delta_{\rm c}(z)$ as the critical density threshold for a
spherical perturbation to collapse by the redshift $z$, and $\delta_{\rm
c1}=\delta_{\rm c}(z_1)$ and $\delta_{\rm c2}=\delta_{\rm c}(z_2)$.  In
the Einstein--de Sitter universe, $\delta_{\rm c}(z)=1.69 (1+z)$.

The probability can be written as 
\begin{eqnarray}
 P(r,M_1,M_2) 
  &=& \frac{1}{\sqrt{2\pi [1-\epsilon^2(r)]}}
      \int^{\infty}_{\nu_{\rm 1c}}
      \exp{\left\{\frac{-[\nu_1-\epsilon(r)\nu_{\rm 2c}]^2}
      {2[1-\epsilon^2(r)]}\right\}}d\nu_1\;, \label{eq:probr}
\end{eqnarray}
where $\nu_1$ and $\nu_2$ are defined by
\begin{equation}
\label{eq:nu}
 \nu_1 \equiv \frac{\delta_1}{\sigma_1} \, , \,
 \nu_2 \equiv \frac{\delta_2}{\sigma_2} \, , \,
 \sigma_1 \equiv \sigma(M_1) \, , \,
 \sigma_2 \equiv \sigma(M_2) \, , \,
 \nu_{\rm 1c} \equiv \frac{\delta_{\rm c1}}{\sigma_1} \, , \,
 \nu_{\rm 2c} \equiv \frac{\delta_{\rm c2}}{\sigma_2} \, , \,
\end{equation}
respectively \citep{yan96}. In equation~(\ref{eq:probr}), $\epsilon(r)$
is defined by
\begin{equation}
 \epsilon(r) \equiv \frac{\sigma_{\rm c}^2(r)}{\sigma_1 \sigma_2}\;,
\end{equation}
where $\sigma_{\rm c}^2(r)$ is the two-point correlation function.  In
equation (\ref{eq:nu}), $\sigma(M)$ is the rms density fluctuation
smoothed over a region of mass $M$.

We can rewrite equation (\ref{eq:probr}) as
\begin{equation}
\label{eq:probr1}
 P(r,M_1,M_2) =\frac{1}{\sqrt{2\pi}}\int_{\beta}^{\infty} 
             e^{-y^2/2} dy \;,
\end{equation}
where
\begin{equation}
\label{eq:beta}
 \beta(r)=\frac{\nu_{\rm 1c}
-\epsilon(r)\nu_{\rm 2c}}{\sqrt{1-\epsilon^2(r)}}
      = \frac{1}{\sqrt{1-\epsilon^2(0)\alpha^2(r)}}
         \frac{\delta_{\rm c1}}{\sigma_1}
\left[1-\frac{\delta_{\rm c2}}{\delta_{\rm c1}}\alpha(r)\right]\:,
\end{equation}
and $\alpha(r)=\epsilon(r)/\epsilon(0)$.  The spatially averaged
conditional probability for $R_{\rm in}<r<R_{\rm out}$ in a precluster
region is defined as
\begin{equation}
\label{eq:prob}
 P(M_1, M_2)=\int_{R_{\rm in}}^{R_{\rm out}} P(r, M_1, M_2) 
4\pi r^2 dr \mbox{\Large /} \int_{R_{\rm in}}^{R_{\rm out}} 4\pi r^2 dr\:.
\end{equation}

The conditional probability that a particle which resides in a object
(`halo') of mass $M_2$ at redshift $z_2$ is contained in a smaller halo
of mass $M_1\sim M_1+d M_1$ at redshift $z_1$ ($z_1>z_2$) is
\begin{equation}
 \label{eq:prob_sps}
P_{\rm SPS}(M_1,t_1|M_2,t_2)d M_1 =
2\left|\frac{\partial P(M, M_2)}{\partial M}\right|_{M=M_1}d M_1 \;,
\end{equation}
where the factor of two is based on the usual Press--Schechter
assumptions.

If we fix $r=0$ in equation~(\ref{eq:probr1}), we can obtain the
conditional probability that a particle which resides in a object of
mass $M_2$ at redshift $z_2$ is contained in a smaller halo of mass
$M_1\sim M_1+d M_1$ at redshift $z_1$ ($z_1>z_2$) in the sense of EPS:
\begin{equation}
 \label{eq:prob_eps}
P_{\rm EPS}(M_1,z_1|M_2,z_2)d M_1 =
2\left|\frac{\partial P(0, M, M_2)}{\partial M}\right|_{M=M_1}d M_1 \;
\end{equation}
\citep{fuj02b}. This means that the probability represented in the EPS
model is that represented at the center of the halo $M_2$ in the SPS
model [$R_{\rm out}\rightarrow 0$ in equation~(\ref{eq:prob_sps})].

We define the typical mass of halos at redshift $z$ that become part of
a larger halo of mass $M_0$ at a later time $z_0 (<z)$ as
\begin{equation}
 \label{eq:m_ave}
 \bar{M}_{\rm SPS}(z|M_0,z_0)
=\frac{\int_{M_{\rm min}}^{M_0} M P_{\rm SPS}(M,z|M_0,z_0)d M}
{\int_{M_{\rm min}}^{M_0} P_{\rm SPS}(M,z|M_0,z_0)d M} \:,
\end{equation}
where $M_{\rm min}$ is the lower cutoff mass. We choose $M_{\rm
min}=10^8\: M_{\odot}$, which corresponds to the mass of dwarf
galaxies. If $R_{\rm in}>0$, equation~(\ref{eq:m_ave}) shows the typical
mass of progenitor halos, excluding the main halo at $r\approx 0$. At
lower redshift, those progenitor halos should be included in the main
halo, and would become subhalos in it. However, at higher redshift, they
should reside around the main halo.

On the other hand, since $P_{\rm EPS}$ corresponds to $P_{\rm SPS}$ when
$R_{\rm out}$ approaches zero, the average mass defined by
\begin{equation}
 \label{eq:m_ave_eps}
 \bar{M}_{\rm EPS}(z|M_0,z_0)
=\frac{\int_{M_{\rm min}}^{M_0} M P_{\rm EPS}(M,z|M_0,z_0)d M}
{\int_{M_{\rm min}}^{M_0} P_{\rm EPS}(M,z|M_0,z_0)d M} \:
\end{equation}
is the expected mass of the main halo that is expected to be at $r\sim
0$. In Fujita (2001a, hereafter Paper~I), we used $\bar{M}_{\rm EPS}$
as the typical mass of progenitors of a cluster. However, the results
should be regarded as those for the main cluster. 

\subsection{Ram-Pressure Stripping}

We adopt the ram-pressure stripping model of Paper~I.  We summerize it
briefly. In the following sections, we often refer to a relatively large
dark halo containing galaxies and gas as a `cluster' and do not
discriminate it from a group or a subcluster.

\subsubsection{Distributions of dark matter and the ICM}
\label{sec:distribution}

The virial radius of a cluster with virial mass $M_{\rm vir}$ is defined
as
\begin{equation}
 \label{eq:r_vir}
r_{\rm vir}=\left[\frac{3\: M_{\rm vir}}
{4\pi \Delta_{\rm c}(z) \rho_{\rm crit}(z)}\right]^{1/3}\:,
\end{equation}
where $\rho_{\rm crit}(z)$ is the critical density of the universe and
$\Delta_{\rm c}(z)$ is the ratio of the average density of the cluster to the
critical density at redshift $z$. The latter is given by
\begin{equation}
\label{eq:Dc_lam}
  \Delta_{\rm c}(z)=18\:\pi^2+82 x-39 x^2\:, 
\end{equation}
for a flat universe with a non-zero cosmological constant
\citep{bry98}. In equation (\ref{eq:Dc_lam}), the parameter $x$ is given
by $x=\Omega(z)-1$. The average density of a cluster, $\rho_{\rm
vir}(z)\equiv\rho_{\rm crit}(z) \Delta_{\rm c}(z)$, is an increasing
function of $z$. In the Einstein--de Sitter Universe, for instance,
$\rho_{\rm vir}(z)\propto (1+z)^3$.

We assume that a cluster is spherically symmetric and the density
distribution of dark matter is
\begin{equation}
 \label{eq:rho_m}
\rho_{\rm m}(r)=\rho_{\rm mv}(r/r_{\rm vir})^{-a}\:,
\end{equation}
where $\rho_{\rm mv}$ and $a$ are constants, and $r$ is the distance
from the cluster center. We choose $a=2.4$, because the slope is
consistent with observations \citep{hor99}. Moreover, the results of
numerical simulations show that the mass distribution in the outer
region of clusters is approximately given by equation (\ref{eq:rho_m})
with $a \sim 2.4$ (Navarro et al. 1996, 1997). We adopt a power-law
rather than the full NFW profile to avoid specifying a particular value
for the concentration parameter and its variation with cluster mass and
formation redshift. We ignore the self-gravity of ICM.

We consider two ICM mass distributions. One follows equation
(\ref{eq:rho_m}), except for the normalization and the core structure,
\begin{equation}
\label{eq:ICM_G}
 \rho_{\rm ICM}(r)=\rho_{\rm ICM, vir}
\frac{[1+(r/r_{\rm c})^2]^{-a/2}}
{[1+(r_{\rm vir}/r_{\rm c})^2]^{-a/2}}\:.
\end{equation}
The ICM mass within the virial radius of a cluster is
\begin{equation}
\label{eq:M_ICM}
 M_{\rm ICM}=\int_{0}^{r_{\rm vir}}4 \pi r^2 \rho_{\rm ICM}(r)dr\:.
\end{equation}
The normalization $\rho_{\rm ICM, vir}$ is determined by the relation
$f_{\rm b}=M_{\rm ICM}/M_{\rm vir}$, where $f_{\rm b}$ is the gas or
baryon fraction of the universe. This distribution corresponds to the
case where the ICM is in pressure equilibrium with the gravity of the
cluster and is not heated by anything other than the gravity. We
introduce the core structure to avoid the divergence of gas density at
$r=0$ and use $r_{\rm c}/r_{\rm vir}=0.1$. We call this distribution the
`non-heated ICM distribution'. We use $f_{\rm b}=0.25 (h/0.5)^{-3/2}$,
where the present value of the Hubble constant is written as
$H_0=100\:h\rm\: km\:s^{-1}\: Mpc^{-1}$. 

We also consider another ICM distribution when ICM is heated
non-gravitationally before cluster formation. At least for nearby
clusters and groups, observations suggest that such non-gravitational
heating did happen \citep{pon99,llo00,mul00,mul03}. Although there is a
debate about whether the heating took place before or after cluster
formation \citep{fuj01b,yam01}, the following results would not be much
different even if ICM is heated after cluster formation
\citep{loe00}. Following \citet{bal99a}, we define the adiabat
$K_0=P/\rho_{\rm g}^{\gamma}$, where $P$ is the gas pressure, $\rho_{\rm
g}$ is its density, and $\gamma\:(=5/3)$ is a constant. If ICM had
already been heated before accreted by a cluster, the ICM fraction of
the cluster is given by
\begin{equation}
\label{eq:f_ICM}
f_{\rm ICM}=\min
\left\{0.040\left(\frac{M_{\rm vir}}{10^{14}\: M_{\odot}}\right)
\left(\frac{K_0}{K_{34}}\right)^{-3/2}
\left[\frac{t(z)}{10^9\rm\: yr}\right],
\:f_{\rm b}\right\}\:,
\end{equation}
where $K_{34}=10^{34}\rm\: erg\:g^{-5/3}\:cm^{2}$ \citep{bal99a}.

The virial temperature of a cluster is given by
\begin{equation}
 \frac{k_{\rm B} T_{\rm vir}}{\mu m_{\rm H}}
=\frac{1}{2}\frac{GM_{\rm vir}}{r_{\rm vir}}\:,
\end{equation}
where $k_{\rm B}$ is the Boltzmann constant, $\mu\;(=0.61)$ is the mean
molecular weight, $m_{\rm H}$ is the hydrogen mass, and $G$ is the
gravitational constant. When the virial temperature of a cluster is much
larger than that of the gas accreted by the cluster, a shock forms near
the virial radius of the cluster \citep{tak98,cav98}. The ICM
temperature inside the shock is related to that of the preshock gas
($T_{\rm p}$) and is approximately given by
\begin{equation}
 T_{\rm ICM} = T_{\rm vir}+\frac{3}{2} T_{\rm p}
\end{equation}
\citep{cav98}. Since we assume that the density profile of dark
matter is given by equation (\ref{eq:rho_m}) with $a=2.4$, the
density profile of ICM is given by
\begin{equation}
\label{eq:ICM_H}
 \rho_{\rm ICM}(r)=\rho_{\rm ICM, vir}
\frac{[1+(r/r_{\rm c})^2]^{-3b/2}}
{[1+(r_{\rm vir}/r_{\rm c})^2]^{-3b/2}}\:, 
\end{equation}
where $b=(2.4/3)T_{\rm vir}/T_{\rm ICM}$ \citep{bah94}. We choose
$3T_{1}/2=0.8$~keV according to Paper~I. The normalization $\rho_{\rm
ICM, vir}$ is determined by the relation $f_{\rm ICM}=M_{\rm ICM}/M_{\rm
vir}$.

When $T_{\rm vir}\lesssim (3/2)T_{\rm p}$, a shock does not form but the
gas accreted by a cluster adiabatically falls into the cluster. The ICM
profile for $r<r_{\rm vir}$ is obtained by solving the equation of
hydrostatic equilibrium and is approximately given by
\begin{equation}
\label{eq:rho_ad}
 \rho_{\rm ICM}(r)=\rho_{\rm ICM, vir}
\left[1+\frac{3}{A}
\ln\left(\frac{r_{\rm vir}}{r}\right)\right]^{3/2}\:,
\end{equation}
where 
\begin{equation}
 A=\frac{15 K_0 \rho_{\rm ICM, vir}^{2/3}}
{8\pi G \rho_{\rm mv,iso}r_{\rm vir}^2}\:
\end{equation}
\citep{bal99a}. Equation~(\ref{eq:rho_ad}) is for the isothermal
dark-matter distribution ($r^{-2}$), which is a little different from
that we adopted ($r^{-2.4}$). We assume equation~(\ref{eq:rho_ad}) as an
approximation because there is no analytical solution for the
dark-matter distribution of $r^{-2.4}$ (Paper~I). The parameter
$\rho_{\rm ICM, vir}$ is determined by the relation $f_{\rm ICM}=M_{\rm
ICM}/M_{\rm vir}$. In section \ref{sec:result}, we use the profile
(\ref{eq:ICM_H}) for $T_{\rm vir}>3T_{1}/2\;(=0.8\rm\; keV)$ and the
profile (\ref{eq:rho_ad}) for $T_{\rm vir}<3T_{1}/2$. We refer to this
ICM distribution as `the heated ICM distribution'. From observations, we
use the value of $K_0=0.37 K_{34}$, which is assumed to be independent
of cluster mass \citep{bal99a}. The heated ICM distribution is flatter
than non-heated ICM distribution, and gives a smaller ICM density at the
cluster center, which has also been demonstrated by numerical
simulations including non-gravitational heating \citep{nav95,bia01}.

\subsubsection{A radially infalling galaxy}
\label{sec:gal}

We consider a radially infalling disk galaxy accreted by a cluster with
dark matter, that is, the first infall to the cluster. The initial
velocity of the model galaxy, $v_{\rm i}$, is given at $r=r_{\rm vir}$, and it
is
\begin{equation}
 \frac{v_{\rm i}^2}{2}=\frac{G M_{\rm vir}}{r_{\rm vir}}
-\frac{G M_{\rm vir}}{r_{\rm ta}}\:,
\end{equation}
where $r_{\rm ta}$ is the turnaround radius of the cluster. Assuming
that $r_{\rm ta}=2\; r_{\rm vir}$ based on the virial theorem, the
initial velocity is
\begin{equation}
 v_{\rm i} = \sqrt{\frac{G M_{\rm vir}}{r_{\rm vir}}}\:.
\end{equation}
The galaxy falls freely toward the cluster center.

As the velocity of the galaxy increases, the ram-pressure from ICM also
increases. The condition of ram-pressure stripping is
\begin{eqnarray}
  \rho_{\rm ICM}v_{\rm rel}^2 
 & >& 2\pi G \Sigma_{\star} \Sigma_{\rm HI} \nonumber\\
 & =& v_{\rm rot}^2 R^{-1} 
  \Sigma_{\rm HI} \label{eq:grav2} \nonumber\\
 & =& 2.1\times 10^{-11}{\rm dyn\: cm^{-2}}
               \left(\frac{v_{\rm rot}}{220\rm\; km\: s^{-1}}\right)^2
               \nonumber\\
 &  &   \times \left(\frac{r_{\rm gal}}{10\rm\; kpc}\right)^{-1}
               \left(\frac{\Sigma_{\rm HI}}
               {8\times 10^{20} 
                   m_{\rm H}\;\rm cm^ {-2}}\right) \label{eq:strip}\:, 
\end{eqnarray}
where $v_{\rm rel}$ is the relative velocity between the galaxy and the
ICM, $\Sigma_{\star}$ is the gravitational surface mass density,
$\Sigma_{\rm HI}$ is the surface density of the H\,{\footnotesize I}
gas, $v_{\rm rot}$ is the rotation velocity, and $r_{\rm gal}$ is the
characteristic radius of the galaxy \citep{gun72,fuj99a}. The original
derivation by Gunn and Gott (1972) assumed a disk mass distribution, but
\citet{aba99} have numerically confirmed that this analytic relation
provides a good approximation even for a more realistic mass
distribution. Note that equation (\ref{eq:strip}) includes the rotation
velocity of the galaxy, and thus it implicitly takes account of the
gravity from the dark-matter halo surrounding the galaxy. From equation
(\ref{eq:strip}) and observations, we can show the dependence of
ram-pressure stripping on the galaxy mass for a given $\Sigma_{\rm
HI}$. \citet{she03} showed that the typical radius and mass of late-type
galaxies have the relation $R\propto M_{\rm gal}^{0.4}$ for $M_{\rm
gal}\gtrsim 10^{10.6}\: M_\odot$ from the Sloan Digital Sky Survey
(SDSS). Since $v_{\rm rot}^2 \propto M_{\rm gal}/R$, we obtain $v_{\rm
rot}^2 R^{-1}\propto M_{\rm gal}^{0.2}$. This means that ram-pressure
stripping is more effective for a smaller galaxy [equation
(\ref{eq:strip})]. We define the cluster radius at which condition
(\ref{eq:strip}) is satisfied for the first time as the stripping
radius, $r_{\rm st}$. Since we assume that the ICM is nearly in pressure
equilibrium for $r<r_{\rm vir}$, the relative velocity, $v_{\rm rel}$,
is equivalent to the velocity of the galaxy relative to the cluster,
$v$, for $r<r_{\rm vir}$. We take account of cluster growth while the
galaxy moves from $r=r_{\rm vir}$ to $r_{\rm st}$ according to Paper~I.

\subsection{Evaporation}

Thermal evaporation of cold gas within galaxies by hot ICM was studied
by Cowie and Songaila (1977), although it seems that the effect of
evaporation has been forgotten for a long time. Basically, the model
presented here is the same as that of Cowie and Songaila (1977).

The energy flux from the hot ICM surrounding a galaxy via thermal
conduction is given by
\begin{equation}
\label{eq:cond1}
 L = -4\pi r^2 \kappa \frac{dT}{dr}\:,
\end{equation}
where $\kappa$ is the conduction rate and is given by $\kappa = \kappa_0
T^{5/2}$. We assume that $\kappa_0 = 5\times 10^{-7}\rm\; erg\;
cm^{-1}\; s^{-1}\; K^{-3.5}$. In general, we do not need to consider the
saturation of thermal conduction if the galaxy is at the center of the
cluster because the ICM density is sufficiently large. In this case,
equation~(\ref{eq:cond1}) is approximated by
\begin{equation}
\label{eq:Lcond}
 |L| \approx 4\pi r_{\rm gal} \kappa_0 T_{\rm ICM}^{7/2}\:.
\end{equation}
Thus, the time scale of the evaporation of cold gas in a galaxy is
written as
\begin{equation}
\label{eq:cond}
 t_{\rm cond}\approx \frac{3}{2}
\frac{k_{\rm B} T_{\rm ICM}}{\mu m_{\rm H}}
\frac{M_{\rm cold}}{|L|} = 
\frac{3}{8\pi}\frac{k_{\rm B}M_{\rm cold}}
{\mu m_{\rm H} r_{\rm gal} \kappa_0 T_{\rm ICM}^{5/2}} \:,
\end{equation}
where $M_{\rm cold}$ is the mass of cold gas in the galaxy.  We define
neutral and molecular gas confined in a galactic disk as cold gas.

On the other hand, if a galaxy is not at the cluster center, the density
of the surrounding ICM would be small and the saturation of thermal
conduction could be important. The saturated flux is given by
\begin{equation}
\label{eq:Lsat}
 |L_{\rm sat}| 
= 4\pi r_{\rm gal}^2\times 0.4 n_{\rm e} k_{\rm B} T_{\rm ICM} 
\left(\frac{2 k_{\rm B}T_{\rm ICM}}{\pi m_{\rm e}}\right)^{1/2}\;,
\end{equation}
where $n_{\rm e}$ is the electron number density and $m_{\rm e}$ is the
electron mass \citep{cow77a}. We define the {\it cluster} radius, $r_{\rm
sat}$, as the one where $|L|=|L_{\rm sat}|$. For $r<r_{\rm sat}$, thermal
conduction is not affected by the saturation. On the other hand, the
flux through thermal conduction is limited by the saturation for
$r>r_{\rm sat}$.

\subsection{Galaxy Mergers}
\label{sec:mod_merger}

If a galaxy is not at the center of a dark halo (here, a dark halo means
a cluster, a subcluster, a group, or its progenitor), it is affected by
dynamical friction and gradually falls toward the halo center. The
galaxy is expected to merge with another galaxy that has already been at
the halo center. In this paper, we treat only galaxy mergers in which
one of the galaxies is at the halo center. Mergers between galaxies that
are not at the halo center (satellite galaxies) are not important unless
the velocity dispersion of the halo is comparable to the rotation
velocities of the galaxies \citep{bin87}.

The time scale of dynamical friction is given by
\begin{equation}
\label{eq:fric}
 t_{\rm fric} \approx \frac{1.17}{\ln \Lambda}
        \frac{r_0 v_{\rm c}}{G M_{\rm gal}}\;, 
\end{equation}
where $\ln \Lambda$ is the Coulomb logarithm, $r_0$ is the initial
position of a galaxy in a halo, $v_{\rm c}$ is the circular velocity of
the halo, and $M_{\rm gal}$ is the mass of the galaxy falling into the
halo center \citep{bin87}. The Coulomb logarithm can be approximated by
$\ln \Lambda \approx (1/2)\ln(1+M_{\rm gal}^2/M_{\rm vir}^2)$
\citep{bin87}. Equation (\ref{eq:fric}) does not depend on the
galaxy types. We set $r_0=r_{\rm h}$, where $r_{\rm h}$ is the half-mass
radius of the halo and it is given by
\begin{equation}
 r_{\rm h} = r_{\rm vir} (1/2)^{1/(3-a)} \:
\end{equation}
for the dark matter distribution we adopt
[equation~(\ref{eq:rho_m})]. We assume that $v_{\rm c}$ is the circular
velocity at $r=r_{\rm h}$.

\section{Results}
\label{sec:result}

As a cosmological model, we consider a cold dark matter model with a
non-zero cosmological constant ($\Lambda$CDM model). The cosmological
parameters are $h=0.7$, $\Omega_0=0.25$, $\lambda_0=0.75$, and
$\sigma_8=0.8$. The mass of a model cluster at $z=0$ is $M_0=1\times
10^{15}\; M_{\odot}$. Among the progenitors, the typical mass of the
main cluster at $z>0$ is given by $M_{\rm vir}=\bar{M}_{\rm
EPS}(z|M_0,0)$. On the other hand, the average mass of the subclusters
is given by $M_{\rm vir}=\bar{M}_{\rm SPS}(z|M_0,0)$. The radius of the
precluster region, $R_0$, is given by $R_0=(3 M_0/4 \pi
\bar{\rho})^{1/3}$, where $\bar{\rho}$ is the current mean mass density
of the universe. Note that $R_0$ is different from the viral radius of
the cluster, $r_{\rm vir}$. We take $R_{\rm in}=0.7 R_0$ and $R_{\rm
out}=R_0$ in equation~(\ref{eq:prob}). If we take $R_{\rm in}=0.5 R_0$
($0.9 R_0$), the average mass of the subclusters becomes larger
(smaller) only by $\sim 20$\%. Thus, the following results are not very
sensitive to the choice of $R_{\rm in}$.

\begin{figure}
  \begin{center}
    \FigureFile(80mm,80mm){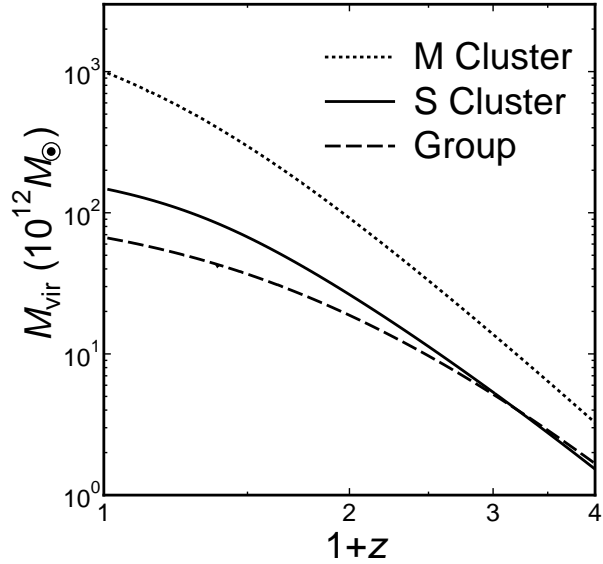}
  \end{center}
  \caption{Mass evolution of a main cluster (dotted
line), a typical subcluster (solid line), and a group (dashed
line).}\label{fig:mass}
\end{figure}

Figure~\ref{fig:mass} shows the evolution of cluster masses. Since
matter is almost uniformly distributed in the very early universe, we
expect that when the mass of the main cluster satisfies the relation
$M_{\rm vir}/M_0 = (R_{\rm in}/R_0)^3$, the subclusters begin to be
included in the main cluster and become the subhalos. For the parameters
we adopted, the subclusters are absorbed by the main cluster at
$z\lesssim 0.4$.

At $z=0.5$, the mass of the main cluster is $3\times 10^{14}\; M_\odot$,
which is close to the masses of well-known clusters observed at $z\sim
0.5$ \citep{sch99}. At $z=0.5$, the average mass of subclusters is
$M_{\rm vir}=6.7\times 10^{13}\; M_{\odot}$ and is on the mass scale of
galaxy groups. In order to compare these subclusters with groups at low
redshift, we also study the evolution of the main component of group
progenitors with the present mass of $M_0=6.7\times 10^{13}\;
M_{\odot}$. The mass at $z>0$ is given by $M_{\rm vir}=\bar{M}_{\rm
EPS}(z|M_0,0)$ and is also shown in figure~\ref{fig:mass}.

\subsection{Ram-Pressure Stripping and Strangulation}
\label{sec:res_ram}

Since we are interested in galaxies at high redshift, we investigate
relatively large galaxies that can be observed in detail. We consider
two model galaxies. From now on, we mostly adopt observed parameters of
the Galaxy for a bigger galaxy and those of M\,33 for a smaller galaxy.
We will show that the results are not sensitive to the galaxy
properties, not only for ram-pressure stripping, but also for
evaporation. In this paper, we focus on the environmental effects on
galaxies; the evolutionary effects of galaxies themselves will be
discussed in a forthcoming paper. The parameters for a bigger model
galaxy are $v_{\rm rot}=220\:\rm km\: s^{-1}$, $r_{\rm gal}=10$ kpc, and
$\Sigma_{\rm HI}=8\times 10^{20}m_{\rm H}\rm\: cm^{-2}$
\citep{spi78}. The parameters for a smaller model galaxy are $v_{\rm
rot}=105\:\rm km\: s^{-1}$, $r_{\rm gal}=5$ kpc, and $\Sigma_{\rm
HI}=14\times 10^{20}m_{\rm H}\rm\: cm^{-2}$ \citep{huc88,sof99}. In
terms of H\,{\footnotesize I} gas, these galaxies are typical for their
luminosities \citep{huc88}.

\begin{figure}
  \begin{center}
    \FigureFile(50mm,50mm){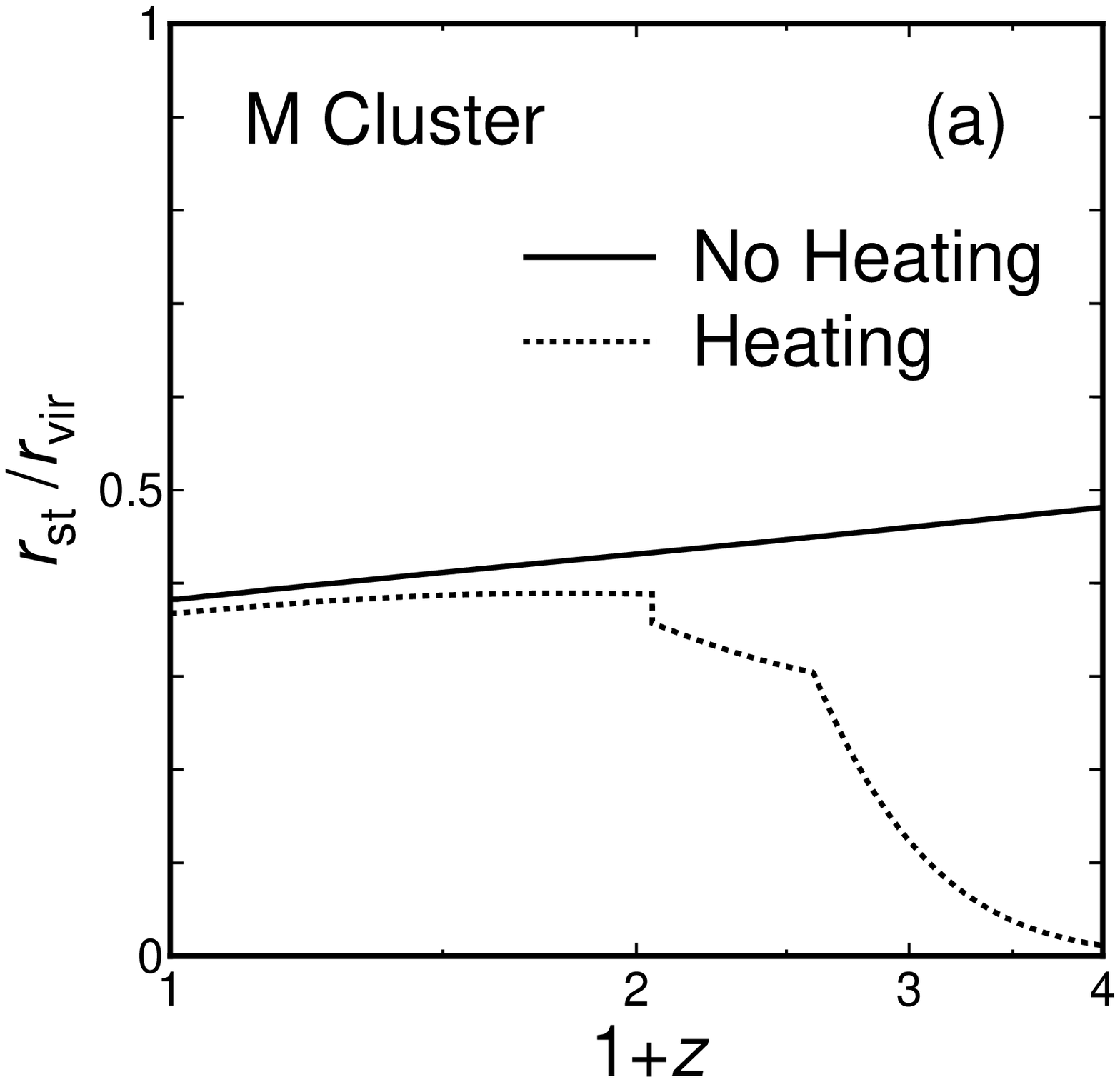}
    \FigureFile(50mm,50mm){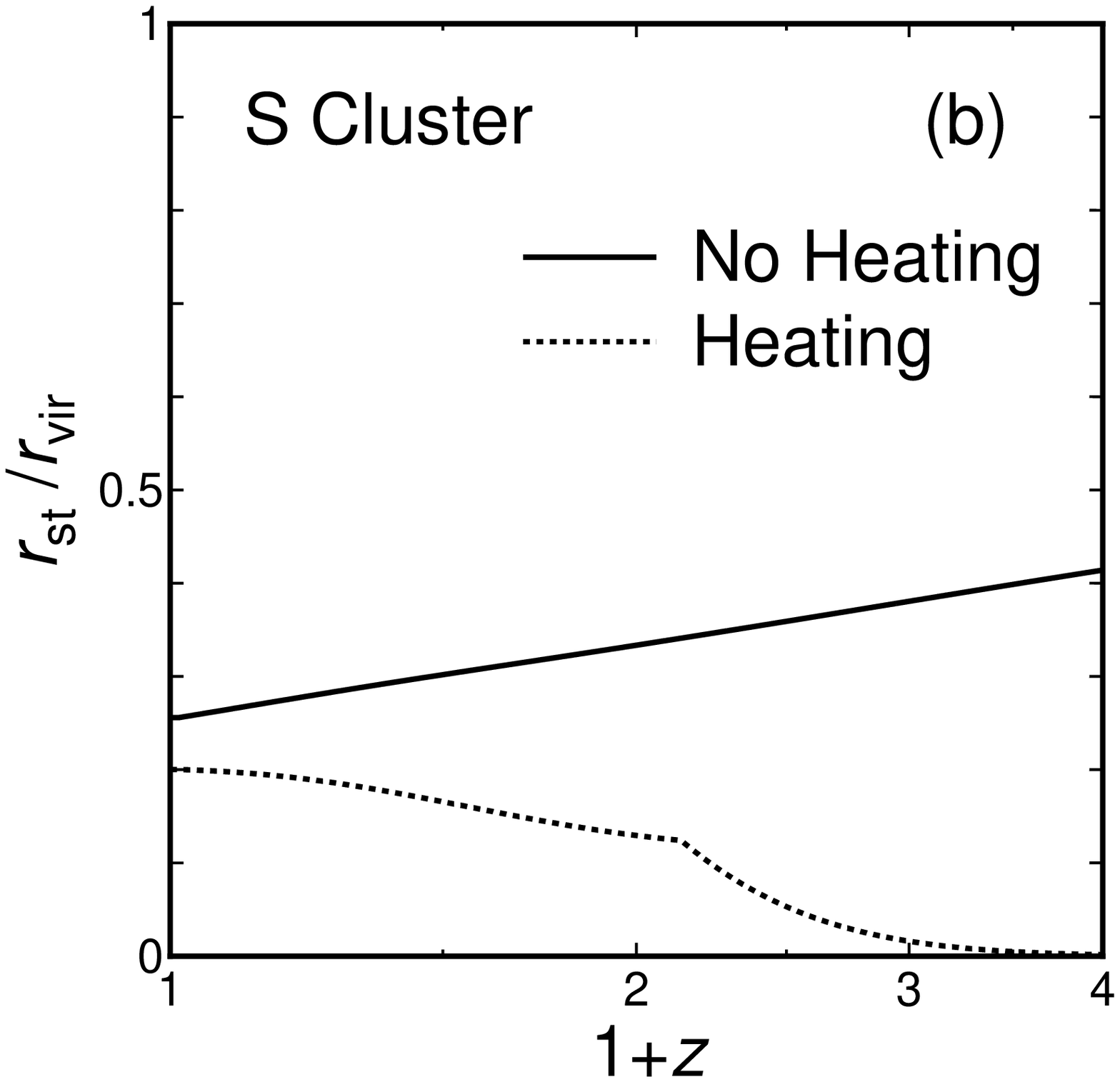}
    \FigureFile(50mm,50mm){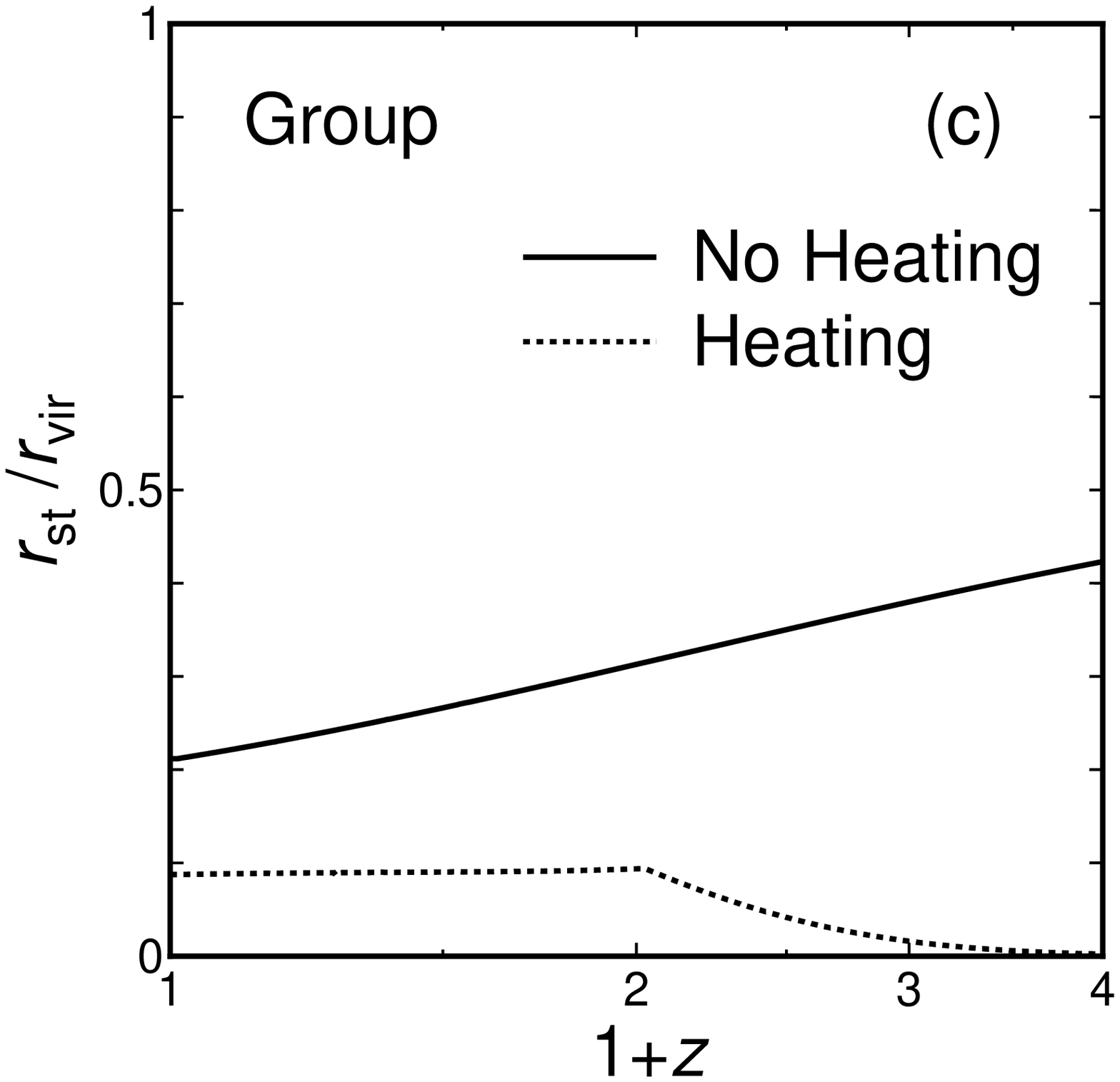}
  \end{center}
  \caption{Evolution of stripping radii, $r_{\rm st}$, normalized by the
virial radii, $r_{\rm vir}$, for a bigger galaxy in a (a) main cluster,
(b) subcluster, and (c) group. The solid lines are the results of the
non-heated ICM model and the dotted lines are those of the heated ICM
model. For $r<r_{\rm st}$, ram-pressure stripping is effective.}
\label{fig:rst}
\end{figure}

\begin{figure}
  \begin{center}
    \FigureFile(50mm,50mm){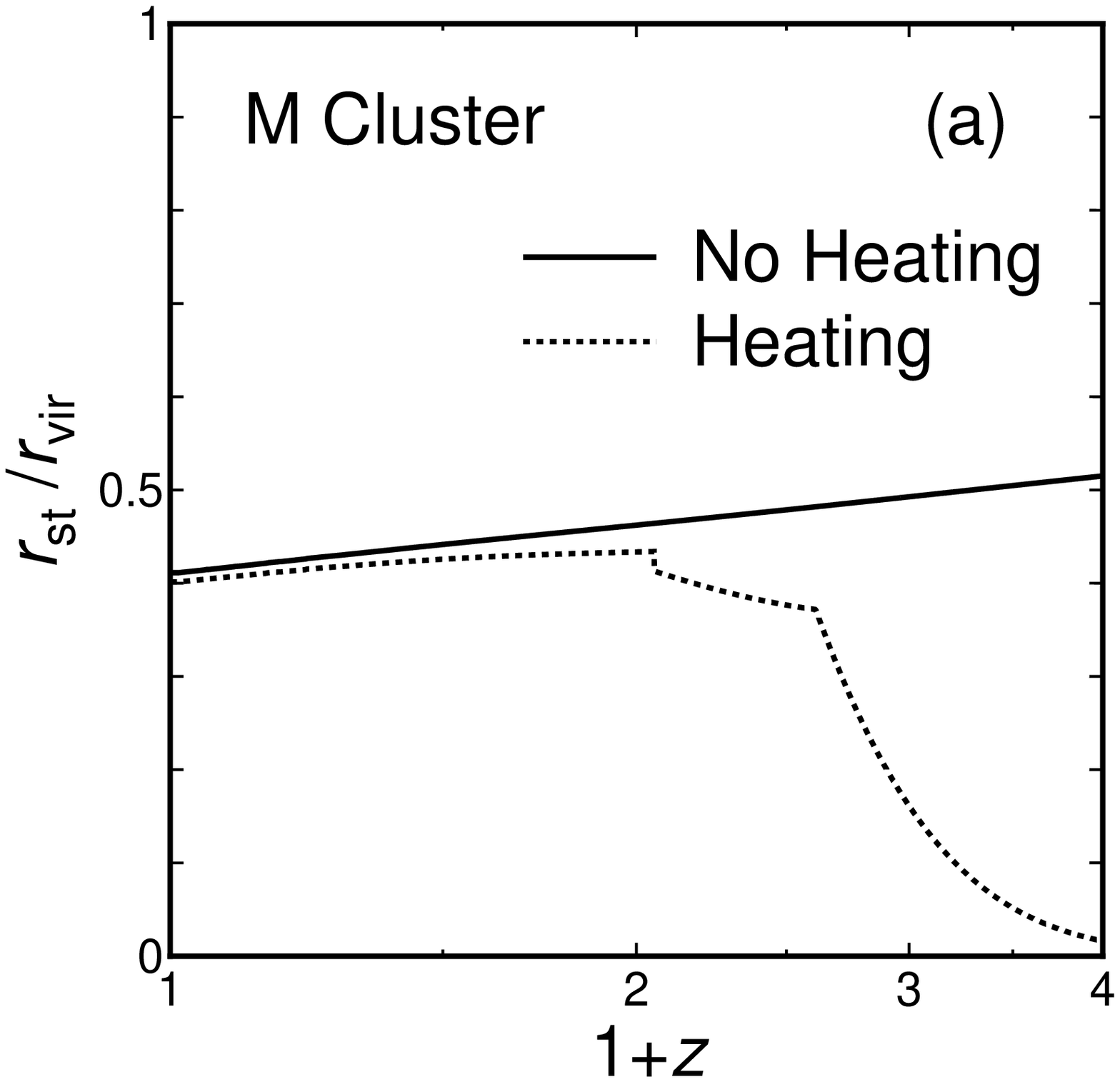}
    \FigureFile(50mm,50mm){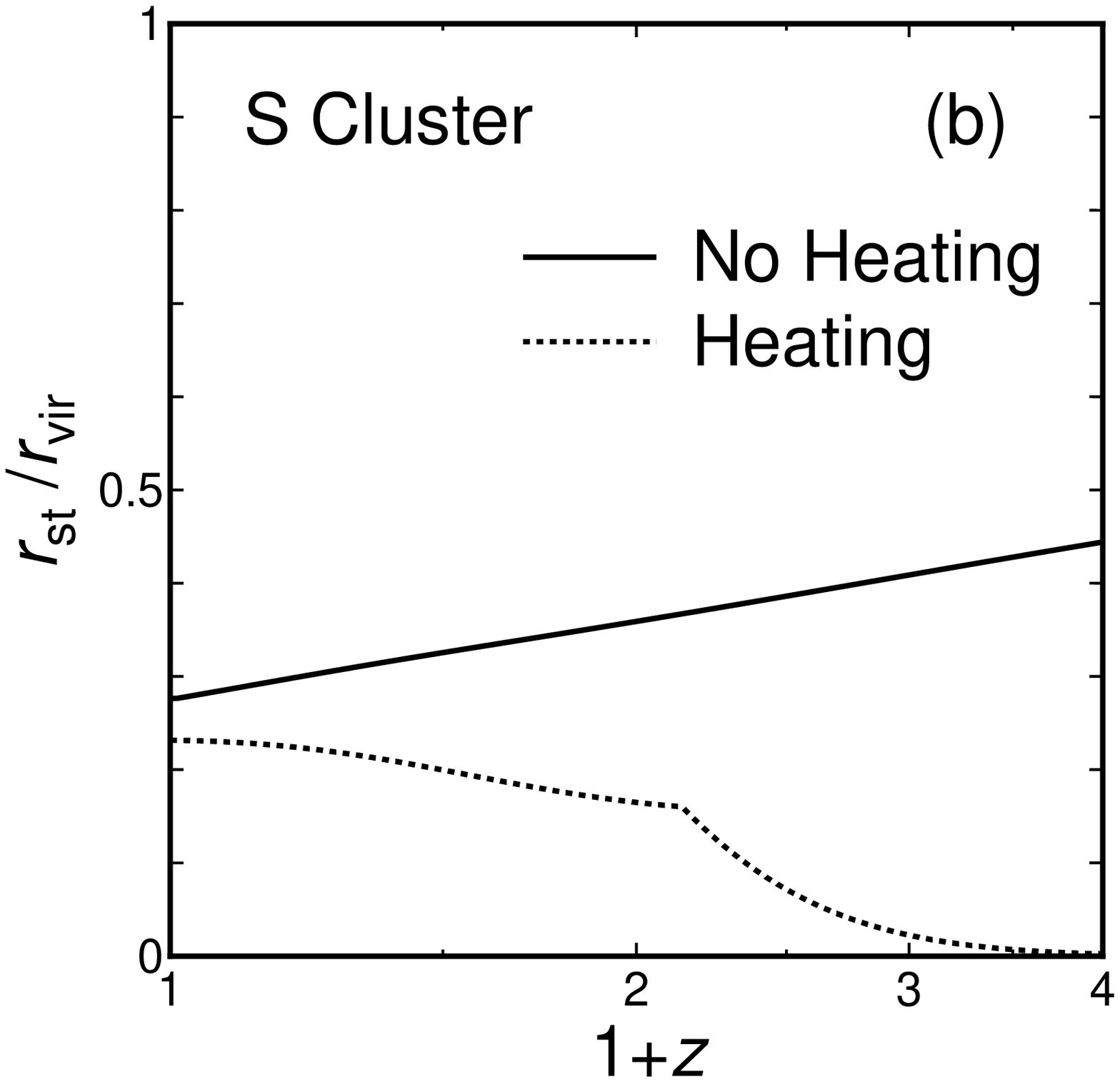}
    \FigureFile(50mm,50mm){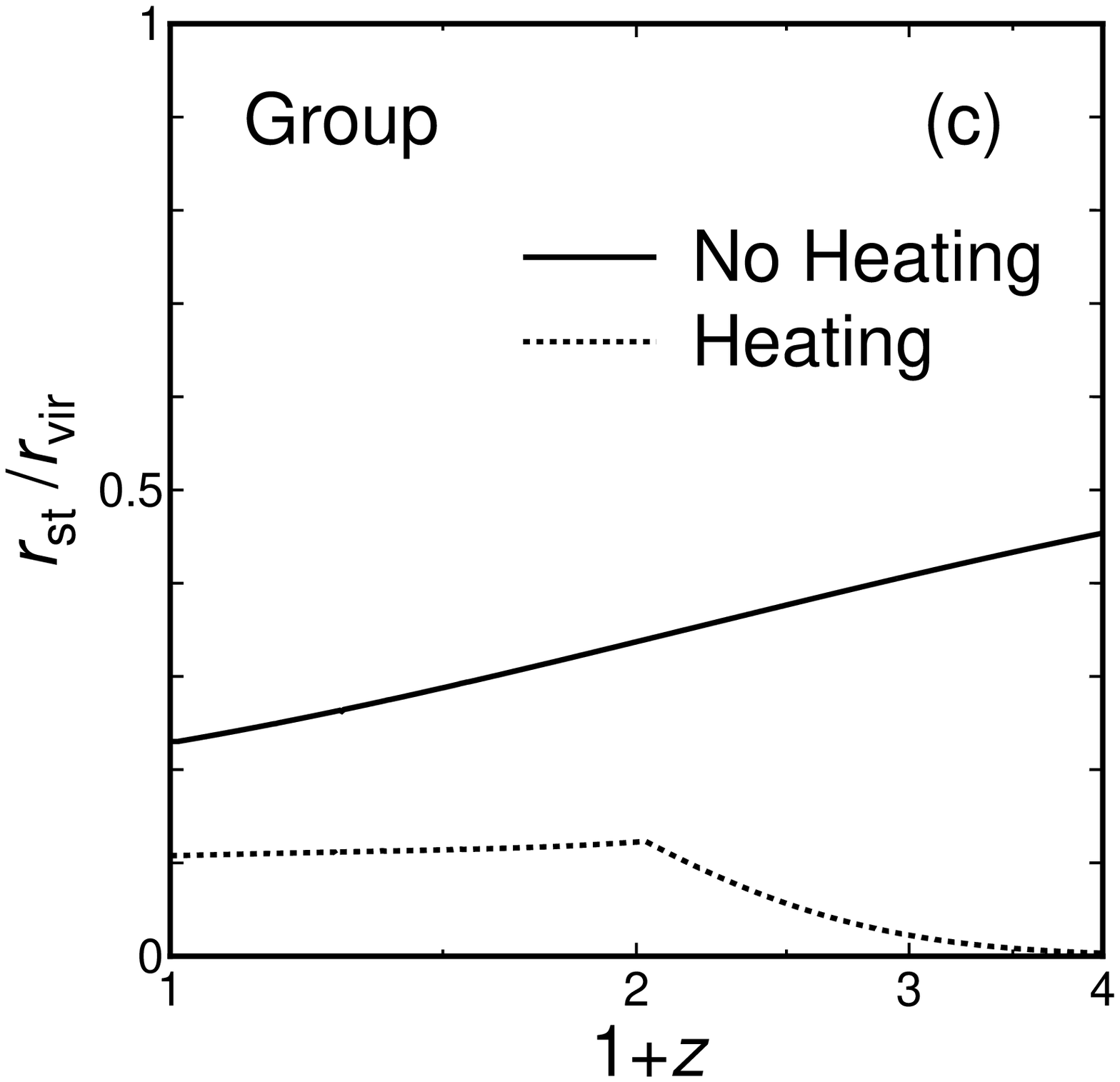}
  \end{center}
  \caption{Same as figure~\ref{fig:rst}, but for a
 smaller galaxy.}
\label{fig:rst_s}
\end{figure}

We show the evolutions of $r_{\rm st}/r_{\rm vir}$ in
figures~\ref{fig:rst} (the bigger galaxy) and~\ref{fig:rst_s} (the
smaller galaxy). Since cold gas in a galaxy is almost instantaneously
stripped by ram-pressure \citep{qui00}, $r_{\rm st}/r_{\rm vir}$ should
be related to the fraction of galaxies affected by ram-pressure
stripping in a cluster. In the models of non-heated ICM, $r_{\rm
st}/r_{\rm vir}$ increases with $z$ for all mass evolution models,
contrary to an expectation that the efficiency of ram-pressure stripping
is low in cluster progenitors at high redshift because their masses are
small and thus galaxy velocities in them are small. The reason for the
increase of $r_{\rm st}/r_{\rm vir}$ at high redshift in our model is
that the average mass density of the progenitors, $\rho_{\rm vir}$, is
large compared to that of clusters or groups at $z=0$. In our model, the
evolution of the average density can be approximated by $\rho_{\rm
vir}\propto (1+z)^3$. If $\rho_{\rm ICM}\propto \rho_{\rm vir}$, the
large mass density leads to large ram-pressure on galaxies although the
masses of the progenitors and the velocities of the galaxies in them
decrease at higher redshift. Taking the subcluster as an example,
$M_{\rm vir}\propto (1+z)^{-3.8}$ approximately
(figure~\ref{fig:mass}). Thus, the virial radius is given by $r_{\rm
vir}\propto (M_{\rm vir}/\rho_{\rm vir})^{1/3}\propto (1+z)^{-2.3}$ and
the typical velocity of galaxies in it is given by $v\propto (M_{\rm
vir}/r_{\rm vir})^{1/2}\propto (1+z)^{-0.77}$. Therefore, the
ram-pressure is represented by $P_{\rm ram}\propto \rho_{\rm
ICM}v^2\propto \rho_{\rm vir}v^2\propto (1+z)^{1.5}$. The mass decrease
rates of the subcluster and the group are smaller than that of the main
cluster (figure~\ref{fig:mass}). This keeps the velocities of galaxies
in the subcluster and the group relatively large even at high
redshift. This is the reason why $r_{\rm st}/r_{\rm vir}$ of the
subcluster and the group increases more rapidly with $z$ than that of
the main cluster. In view of the large $r_{\rm st}/r_{\rm vir}$, it is
likely that galaxies in subclusters around the main cluster are affected
by ram-pressure stripping before they fall into the main cluster, if the
ICM of the subclusters has not been heated non-gravitationally.

Recently, \citet{tor03} studied orbital properties of galaxies in
massive clusters by numerical simulations. They showed that the typical
pericentric radius of a galaxy for the first passage is about $0.3\:
r_{\rm vir}$. This means that most galaxies should be affected by
ram-pressure stripping in massive clusters if non-gravitational heating
can be ignored (figures~\ref{fig:rst}a and~\ref{fig:rst_s}a). On the
other hand, it is not certain that the results of \citet{tor03} can be
applied to the subclusters or groups with smaller masses. If they can,
it is shown that ram-pressure stripping takes place for most galaxies
for $z\gtrsim 0$--$0.5$ in the subcluster (figures~\ref{fig:rst}b
and~\ref{fig:rst_s}b) and $z\gtrsim 0.5$--$1$ in the group
(figures~\ref{fig:rst}c and~\ref{fig:rst_s}c).

The differences between figures~\ref{fig:rst} and~\ref{fig:rst_s} are
small. This is because $P_{\rm ram}$ changes very rapidly as a galaxy
falls toward the cluster center. Thus, although the thresholds of
ram-pressure stripping are different between the bigger and smaller
galaxies [Equation~(\ref{eq:strip})], $r_{\rm st}/r_{\rm vir}$ is not
significantly affected by the difference.

We can compare ram-pressure stripping in subclusters at high redshift
and that in groups at low redshift. The mass of the subcluster at
$z=0.5$ equals that of the group at $z=0$. However, $r_{\rm st}/r_{\rm
vir}$ of the subcluster at $z=0.5$ is larger than that of the group at
$z=0$ (figures~\ref{fig:rst}b and~\ref{fig:rst}c, or
figures~\ref{fig:rst_s}b and~\ref{fig:rst_s}c) because of the large mass
density of the subcluster. This means that ram-pressure stripping is
more effective in the subcluster at $z=0.5$ than in the group at $z=0$,
although they have the same mass. At higher redshift, the efficiency of
ram-pressure stripping increases further.

For a given redshift, $r_{\rm st}/r_{\rm vir}$ of the main cluster is
larger than those of the subcluster and the group (figures~\ref{fig:rst}
and~\ref{fig:rst_s}). This means that ram-pressure stripping is more
effective in more massive clusters. This is because the typical velocity
of galaxies is larger for more massive clusters (Paper~I).

In the model of heated ICM, $r_{\rm st}/r_{\rm vir}$ decreases at high
redshift (figures~\ref{fig:rst} and~\ref{fig:rst_s}). The changes of the
slope are due to the shift in the ICM distribution from
equation~(\ref{eq:ICM_H}) to (\ref{eq:rho_ad}) and the shift in the ICM
fraction from $f_{\rm ICM}=f_{\rm b}$ to $f_{\rm ICM}<f_{\rm b}$
[equation~(\ref{eq:f_ICM})] toward high redshift. Taking the heated ICM
model for the main cluster as an example (figures~\ref{fig:rst}a
and~\ref{fig:rst_s}a), the shift in the ICM distribution occurs at
$z=1.0$ and the shift in $f_{\rm ICM}$ occurs at $z=1.6$.  For a given
redshift, the ratios $r_{\rm st}/r_{\rm vir}$ in the heated ICM model
are smaller than those in the non-heated ICM model because the ICM
densities are smaller. The difference is more significant for the
subcluster and the group than for the main cluster. This is because the
ICM distribution is more flattened in less massive clusters, and
especially the ICM densities in the central regions decrease
dramatically through non-gravitational heating
(subsection~\ref{sec:distribution}). For $z\gtrsim 2$--$3$, $r_{\rm
st}/r_{\rm vir}$ is very small for all mass evolution models because the
masses are very small in this redshift range
(figure~\ref{fig:mass}). Thus, we predict that ram-pressure stripping
does not occur in the redshift range. On the other hand, the ratio
$r_{\rm st}/r_{\rm vir}$ in the heated ICM model is almost the same as
that in the non-heated ICM model for the main cluster at low redshift
(figures~\ref{fig:rst}a and~\ref{fig:rst_s}a). The ratio $r_{\rm
st}/r_{\rm vir}$ for the subcluster at $z=0.5$ is larger than that for
the group at $z=0$ (figures~\ref{fig:rst}b and~\ref{fig:rst}c or
figures~\ref{fig:rst_s}b and~\ref{fig:rst_s}c). This appears to show
that ram-pressure stripping is {\it relatively} effective in subclusters
at $z\sim 0.5$ even when non-gravitational heating has
occurred. However, in the subclusters, the {\it absolute} fraction of
galaxies affected by ram-pressure stripping may not be large. Assuming
that a subcluster is located near the virial radius of the main cluster,
the tidal acceleration from the main cluster affects the subcluster and
the galaxies in it. It is given by
\begin{equation}
\label{eq:gacc}
 a_{\rm t} \approx \frac{G M_{\rm vir,main}}{r_{\rm vir,main}^2} \;,
\end{equation}
where $M_{\rm vir,main}$ and $r_{\rm vir,main}$ are the virial mass and
radius of the main cluster, respectively. We assume that a galaxy falls
towards the center of the subcluster on an exactly radial orbit at the
virial radius of the subcluster, $r_{\rm vir,sub}$. When the galaxy
reaches the center of the subcluster, the orbit of the galaxy in the
subcluster would be shifted by $\Delta r \approx a_{\rm t} t_{\rm
cent}^2/2$, where $t_{\rm cent}$ is the time in which the galaxy moves
from $r=r_{\rm vir,sub}$ to the center of the subcluster. In our models,
for the main cluster and the subcluster at $z=0.5$, the shift is $\Delta
r/r_{\rm vir,sub}\approx 0.2$, although the estimation is very
rough. Comparing it with figures~\ref{fig:rst}b or~\ref{fig:rst_s}b, it
is shown that the galaxy can dodge the central region of the subcluster
where ram-pressure stripping is effective ($r<r_{\rm st}$) if the ICM
has non-gravitationally been heated. We note that if the results of
\citet{tor03} claiming that pericentric radii of galaxies are $\sim
0.3\: r_{\rm vir}$ can be applied to the subcluster and group regardless
of the redshift, ram-pressure stripping is ineffective for most galaxies
in the subcluster and group regardless of redshift if the ICM has
non-gravitationally been heated.

In summary, figures~\ref{fig:rst} and~\ref{fig:rst_s} show that if ICM
(or the gas accreted by a cluster later on) is heated
non-gravitationally at a redshift well over one, ram-pressure stripping
does not occur in cluster progenitors including both main clusters and
subclusters at $z\gtrsim 2$--$3$. Moreover, for subclusters around the
main cluster, tidal acceleration may prevent galaxies from being
affected by ram-pressure stripping even at lower redshift. On the other
hand, if the ICM has not been heated non-gravitationally until $z\sim
0$, ram-pressure stripping occurs even at $z\sim 3$ in cluster
progenitors.

In the above discussion, we assumed that cold gas remains in a galactic
disk until the condition of ram-pressure stripping
[equation~(\ref{eq:strip})] is satisfied. This is the case if the gas
circulation in a galaxy is confined to the cold gas and stars. However,
if warm gas filling in the galactic halo also joins the circulation and
is indispensable for the supply of the cold gas, the ram-pressure
stripping of the warm gas would be important for the evolution of the
galaxy. The low-density warm gas would be stripped before the condition
of the ram-pressure stripping of the high-density cold gas is met. After
the warm gas is stripped, the cold gas will dissappear in the time scale
of star formation of the galaxy, $\tau_{\star}$, because the gas supply
from the warm gas phase is cut and the cold gas in the galaxy is
gradually consumed by star formation in the galaxy (strangulation). The
time scale of star formation is expected to be $\tau_\star>1$~Gyr
\citep{oka03}. After the cold gas is consumed completely, the star
formation of the galaxy dies out. Note that for a radially infalling
galaxy in an actual cluster, rapidly increasing ram-pressure decreases
the star-formation time scale and increases the star-formation
efficiency of molecular gas in the galaxy significantly
\citep{elm97,fuj99a}. These accelerate the consumption of the cold
gas. However, if stripping is ignored, the time scale of the gas
consumption, $\tau_{\star}$, is not likely to be smaller than Gyr even
if the gas compression by ram-pressure is considered (figures~3 and~4 in
\cite{fuj99a}). On the other hand, once the condition of ram-pressure
stripping is satisfied, the cold gas of a galaxy is abruptly removed and
the star-formation activity of the galaxy is turned off in a very short
time because there is no source of stars any longer ($\sim 10^8$~yr;
\cite{fuj99a,qui00}). Thus, when we assume that for a galaxy falling
into a cluster from the maximum expansion radius of the cluster, the
time scale of the warm gas stripping, $t_{\rm wst}$, is limited to the
time in which the galaxy moves from $r=r_{\rm vir}$ to $r=r_{\rm
st}$. Therefore, for the galaxy, the maximum time scale of the decrease
of the star-formation rate when the contribution of warm gas to the gas
circulation in the galaxy is considered is represented by $t_{\rm
stra}=\min(\tau_\star, t_{\rm wst})$.

\begin{figure}
  \begin{center}
    \FigureFile(80mm,80mm){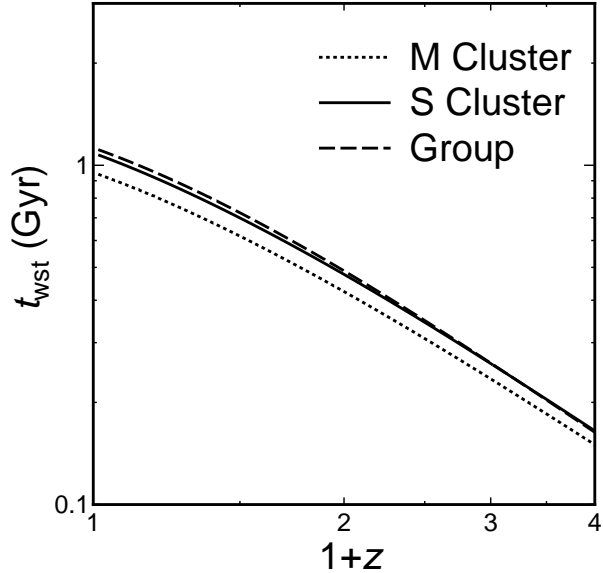}
  \end{center}
  \caption{Maximum time scale allowed for strangulation or evaporation
 for a main cluster (dotted line), a subcluster (solid line), and a
 group (dashed line).} \label{fig:twst}
\end{figure}

Figure~\ref{fig:twst} shows the redshift dependence of $t_{\rm wst}$ for
the non-heated ICM model. Since $t_{\rm wst}\lesssim 1$~Gyr, we expect
that $t_{\rm wst}\lesssim \tau_\star$ and $t_{\rm stra}=t_{\rm wst}$,
especially when $z$ is large. This means that for galaxies infalling
from the outside of a cluster or its progenitors, ram-pressure stripping
of the cold gas cannot be avoided if their orbits are close to radial
ones. The maximum time scale of the decrease of the star-formation rates
($t_{\rm stra}<1$~Gyr) will be useful to be compared with
observations. Okamoto and Nagashima (2003) conducted $N$-body
simulations combined with a semi-analytic model of galaxy
formation. They concluded that ram-pressure stripping does not affect
the galaxy evolution in clusters. However, it may be too early to
conclude as they did because they did not consider the ram-pressure
stripping in subclusters. Calculations including the effect of
ram-pressure stripping in subclusters would be interesting.

For the heated ICM model, $t_{\rm wst}$ is not much different from that
for the non-heated ICM model, because it takes little time for a galaxy
to cross the central region of a cluster and thus the difference of
$r_{\rm st}$ does not much affect $t_{\rm wst}$. Since $r_{\rm
st}/r_{\rm vir}$ in the heated ICM model is very small for the main
cluster at $z\gtrsim 3$ and for the group at $z\gtrsim 2$
(figures~\ref{fig:rst} and~\ref{fig:rst_s}), ram-pressure stripping of
cold gas may not occur at these redshifts. For the subcluster,
ram-pressure stripping may not take place even at lower redshift,
because the tidal force from the main cluster may have galaxies elude
the central region of the subcluster where ram-pressure stripping is
effective [equation~(\ref{eq:gacc})]. In these cases, only warm gas
stripping or strangulation may occur. Here, we would like to emphasize
that {\it even if future observations such as the search for K$+$A
galaxies show that ram-pressure stripping does not take place in
subclusters or main clusters at high redshift, it is not an obvious
result.}  The small masses of the subclusters or main clusters are not
the only reason. It requires additional reasons such as
non-gravitational heating and/or tidal acceleration.

\subsection{Evaporation}

\begin{figure}
  \begin{center}
    \FigureFile(50mm,50mm){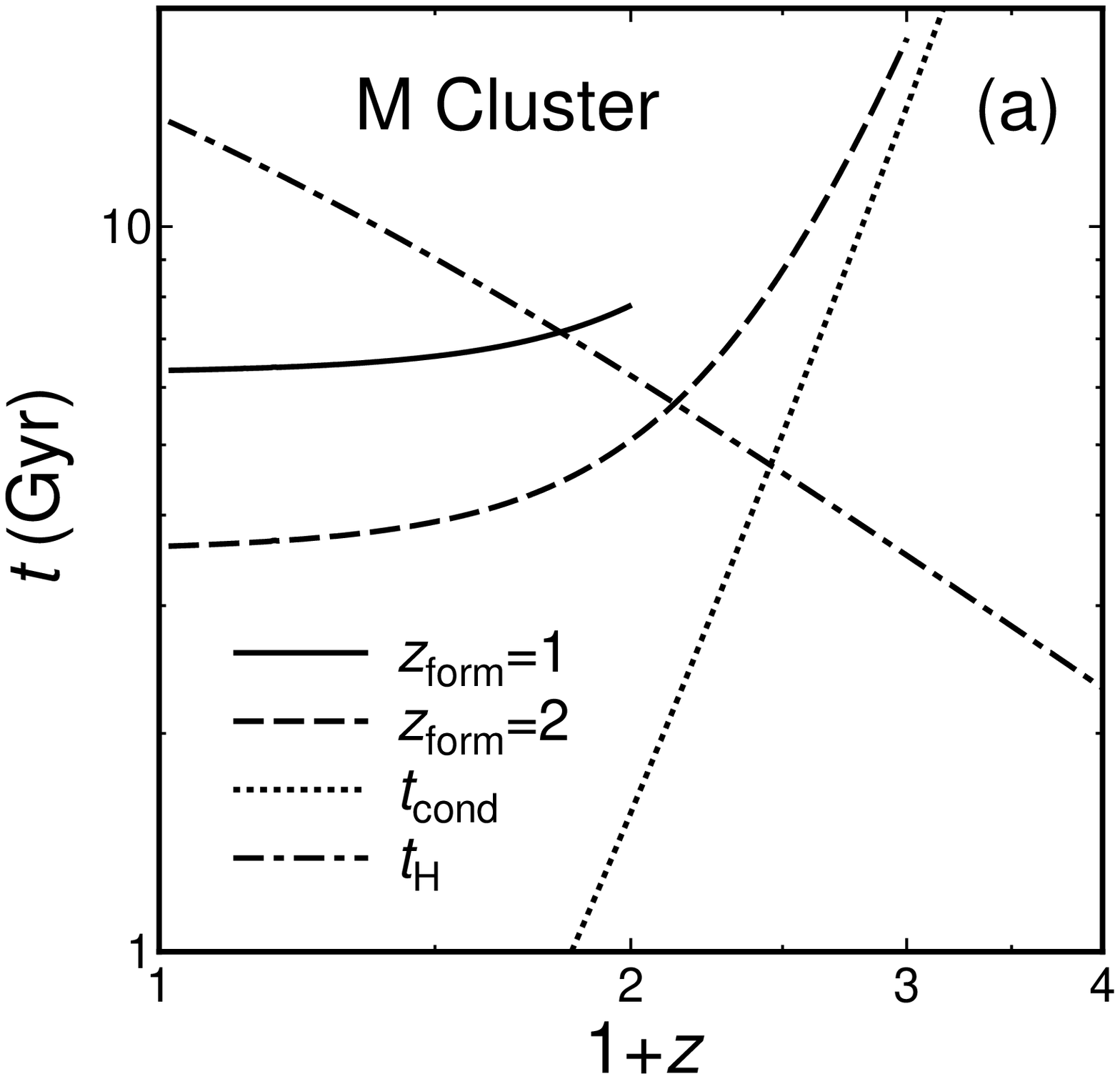}
    \FigureFile(50mm,50mm){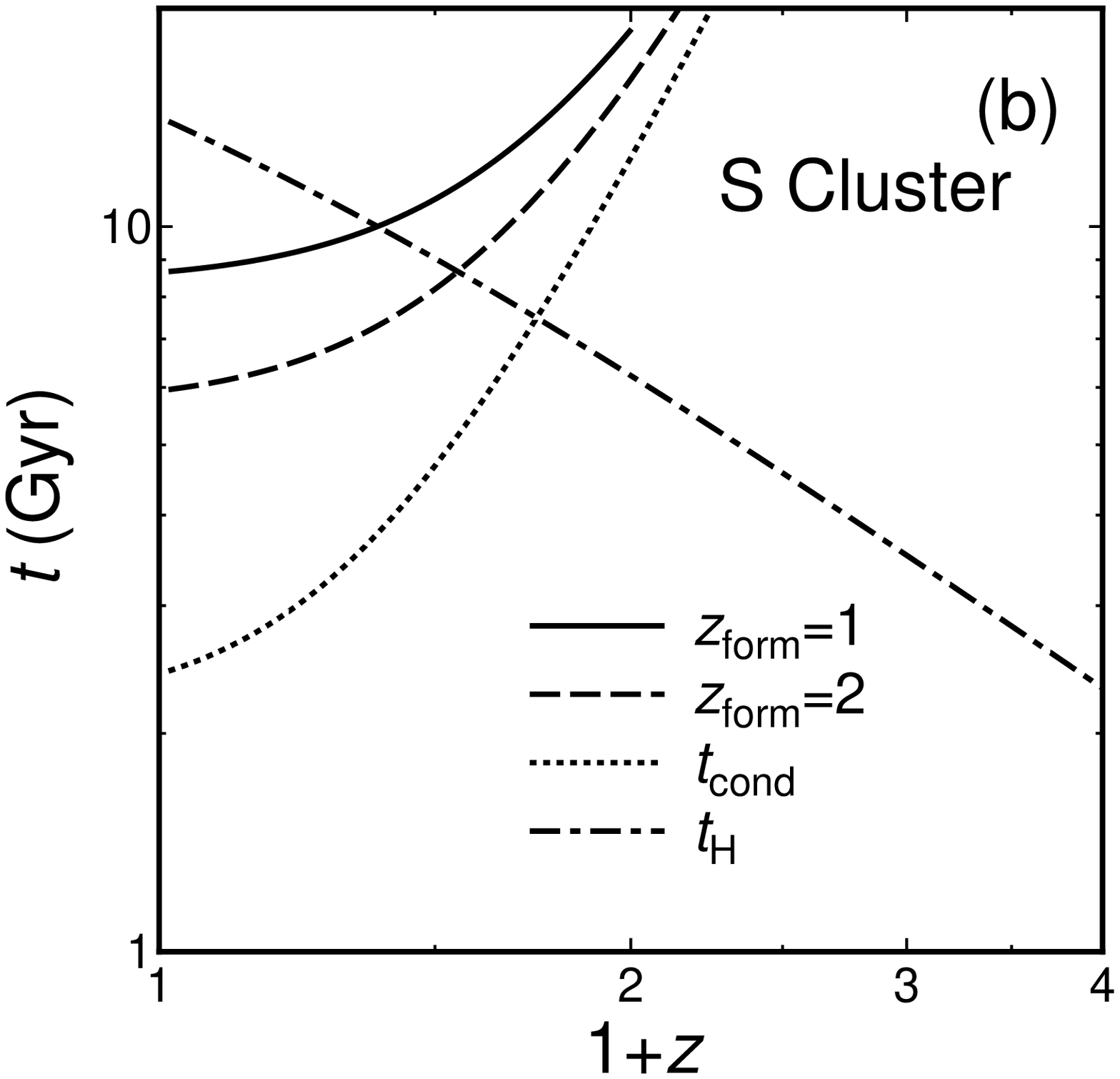}
    \FigureFile(50mm,50mm){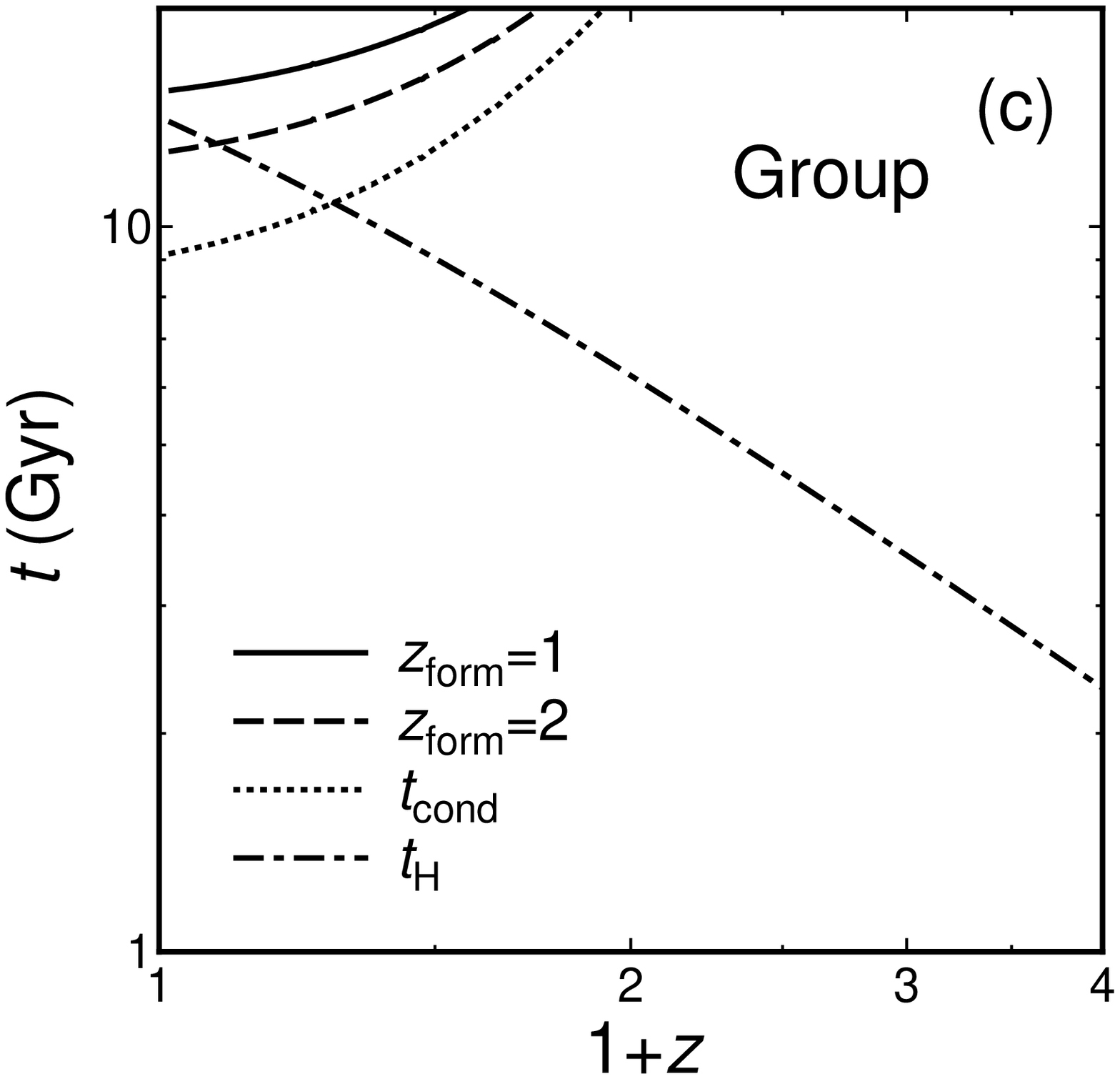}
  \end{center}
  \caption{Time scales of thermal conduction for a
 bigger galaxy in a (a) main cluster, (b) subcluster, and (c)
 group. The solid and dashed lines show $t_{\rm cond}'=t_{\rm cond}+t_{\rm
 form}$ when $z_{\rm form}=1$ and~2, respectively. The dotted and
 dot-dashed lines show $t_{\rm cond}$ and the Hubble time $t_{\rm H}$,
 respectively.  For $t_{\rm cond}'\lesssim t_{\rm H}$, thermal
 conduction is effective.} \label{fig:tcond}
\end{figure}

\begin{figure}
  \begin{center}
    \FigureFile(50mm,50mm){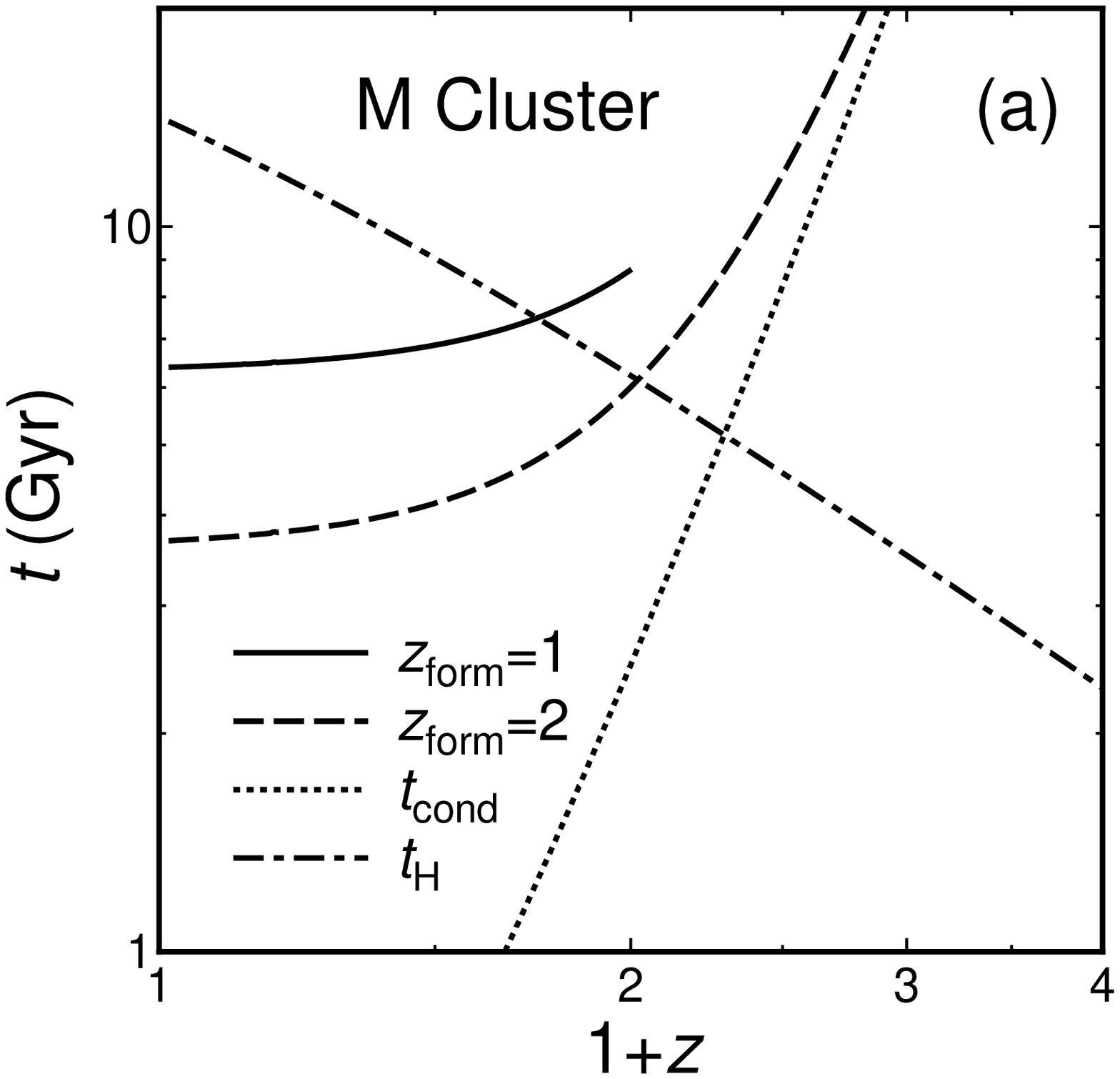}
    \FigureFile(50mm,50mm){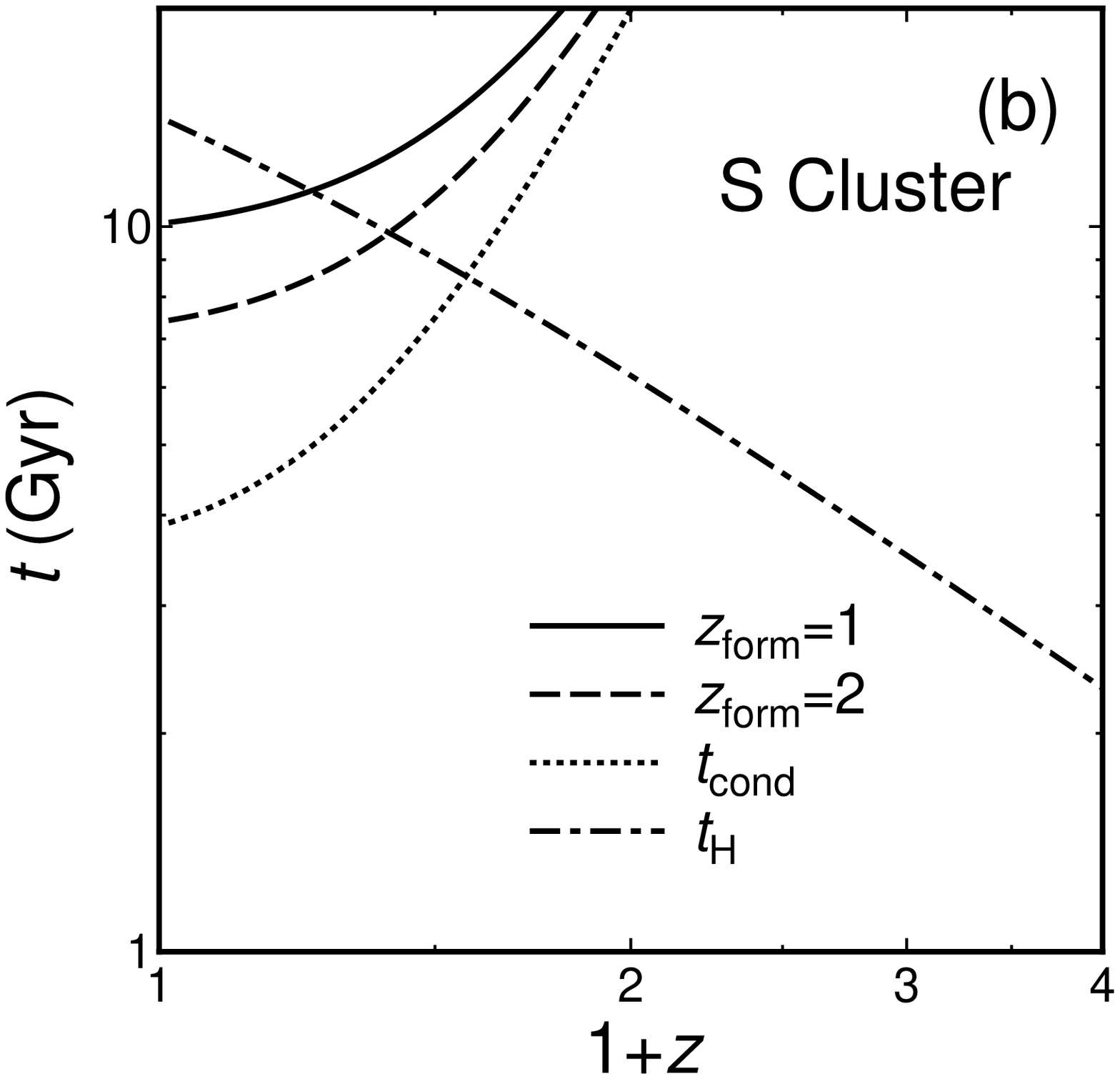}
    \FigureFile(50mm,50mm){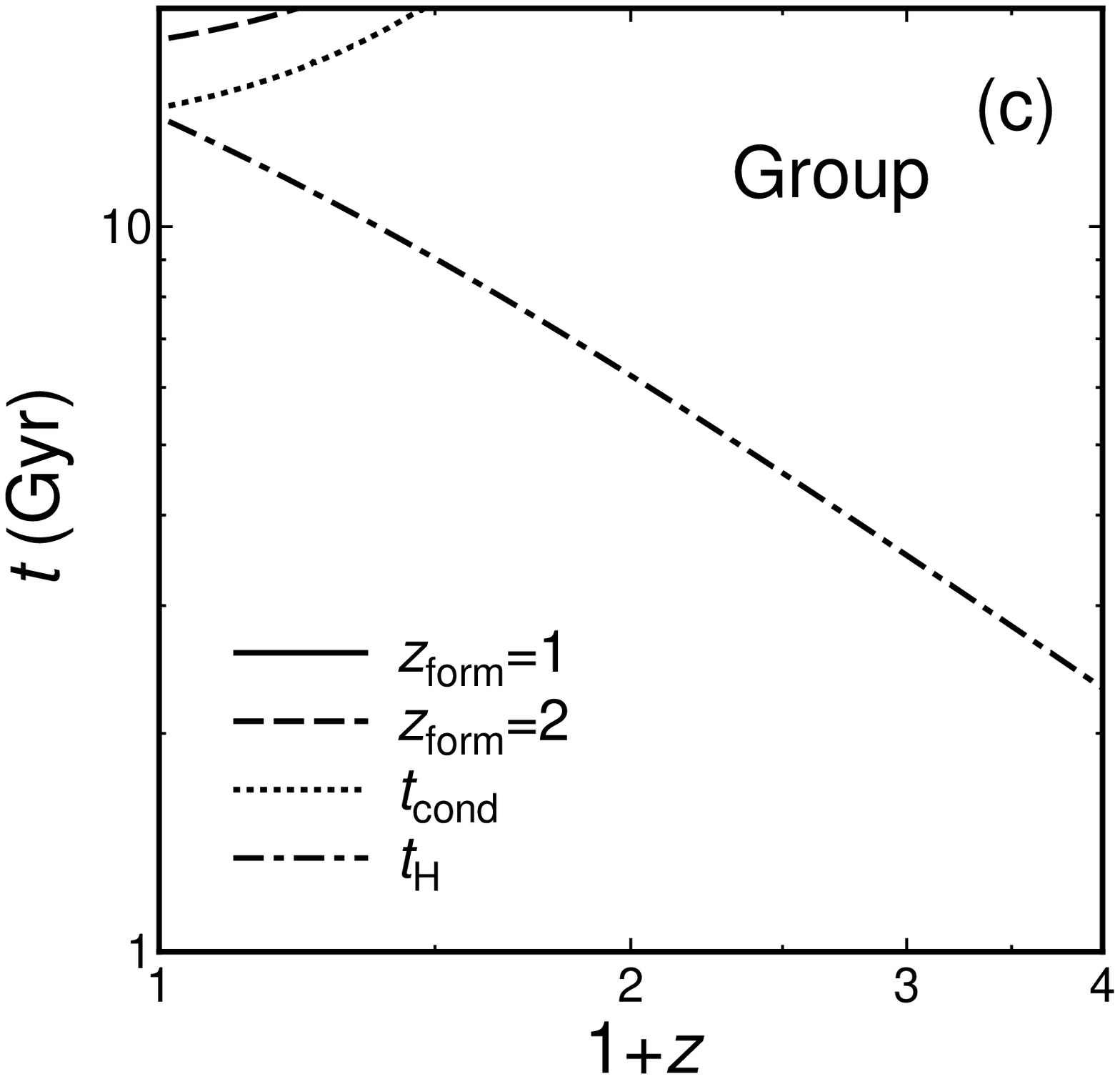}
  \end{center}
  \caption{Same as figure~\ref{fig:tcond}, but for a
 smaller galaxy.}\label{fig:tcond_s}
\end{figure}

A galaxy at the center of a cluster does not move relative to the
cluster. Thus, it is not affected by ram-pressure stripping,
strangulation, and dynamical friction, although it may be the target of
other galaxies falling through dynamical friction. However, even if the
central galaxy is not affected by these mechanisms, it would be heated
by the surrounding hot ICM via thermal conduction. The time scale in
which cold gas in a galaxy is evaporated is represented by $t_{\rm
cond}$ [equation~(\ref{eq:cond})] and is presented in
figures~\ref{fig:tcond} and~\ref{fig:tcond_s} for a bigger and a smaller
galaxy, respectively. In these calculations, we assumed that $r_{\rm
gal}=10$~kpc and $M_{\rm cold}=5\times 10^9\; M_\odot$ for the bigger
galaxy, and $r_{\rm gal}=5$~kpc and $M_{\rm cold}=4\times 10^9\;
M_\odot$ for the smaller galaxy. We also assumed that $T_{\rm
ICM}=T_{\rm vir}$. We ignore the gas supply from the ICM to the central
galaxy through a cooling flow. Figures~\ref{fig:tcond}
and~\ref{fig:tcond_s} show that $t_{\rm cond}$ rapidly decreases as $z$
decreases. This is because $t_{\rm cond}$ is very sensitive to $ T_{\rm
ICM}$ [equation~(\ref{eq:cond})] and $T_{\rm ICM}$ increases as the mass
of a cluster (or a subcluster or a group) increases. Because of the
sensitivity, the results do not much depend on the deference of galaxy
characters (figures~\ref{fig:tcond} and~\ref{fig:tcond_s}). In these
figures, the Hubble time, $t_{\rm H}$, is also presented. At low
redshift, figures~\ref{fig:tcond} and~\ref{fig:tcond_s} show that
$t_{\rm cond}\ll t_{\rm H}$ in some cases, and thus the cold gas in a
galaxy is evaporated soon after the galaxy forms in the cluster or its
progenitors. Therefore, we also show $t_{\rm cond}'=t_{\rm cond}+t_{\rm
form}$ in the figures, where $t_{\rm form}$ is the Hubble time when the
galaxy forms at $z=z_{\rm form}$. If $t_{\rm cond}'<t_{\rm H}$ at a
given redshift, the cold gas in a galaxy has been evaporated by the
redshift. Note that $t_{\rm cond}$ corresponds to $t_{\rm cond}'$ when
$z_{\rm form}=\infty$. For the bigger galaxy in the subcluster, the
evaporation is effective at $z\lesssim 0.4$ and~0.6 if $z_{\rm form}=1$
and~2, respectively (figure~\ref{fig:tcond}b). For the smaller galaxy in
the subcluster, the evaporation is effective at $z\lesssim 0.3$ and~0.4
if $z_{\rm form}=1$ and~2, respectively
(figure~\ref{fig:tcond_s}b). This means that the evaporation may be
developing in the subclusters around the main clusters observed at
$z\sim 0.5$. However, since $t_{\rm cond}\gtrsim 2$~Gyr at the redshift
when the condition $t_{\rm cond}'\lesssim t_{\rm H}$ is satisfied
(figures~\ref{fig:tcond}b and~\ref{fig:tcond_s}b), the truncation of
star formation in the galaxy proceeds much more slowly than those by
ram-pressure stripping and strangulation accompanied by ram-pressure
stripping at the end. If gas consumption by star formation is
considered, the truncation by evaporation would take a less time.

For the main cluster, evaporation begins to be effective at higher
redshift. Figure~\ref{fig:tcond}a and~\ref{fig:tcond_s}a indicate that
cold gas in galaxies should have been evaporated in massive clusters at
$z\sim 1$, although galaxies at the centers of massive clusters are not
disk galaxies but mostly cD galaxies with little cold gas. For the
group, evaporation becomes effective at lower redshift
(figures~\ref{fig:tcond}c and~\ref{fig:tcond_s}c). Thus, it may be
undergoing at $z\sim 0$ and the situation may be similar to that of the
subclusters observed at $z\sim 0.5$.

\begin{figure}
  \begin{center}
    \FigureFile(50mm,50mm){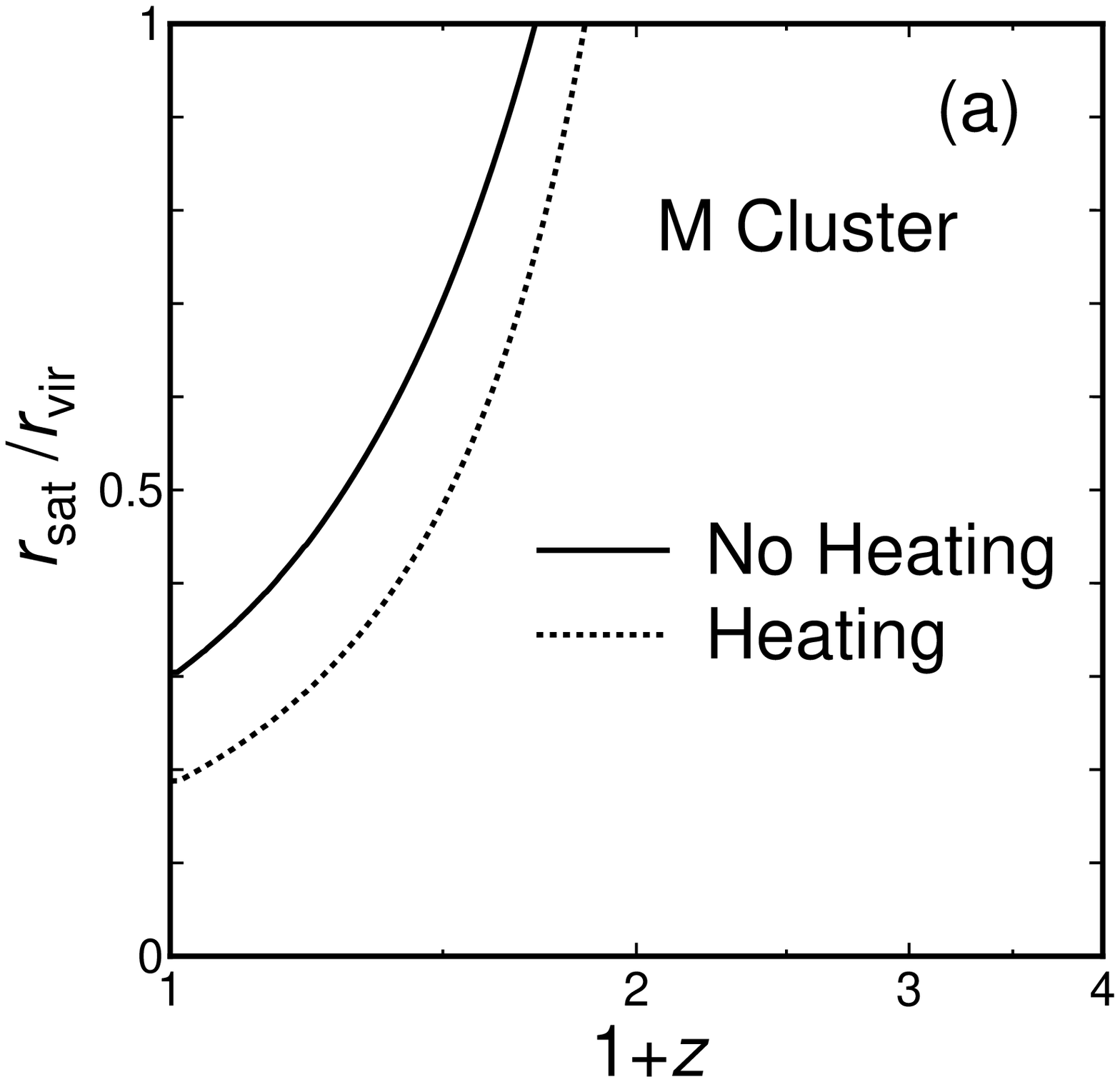}
    \FigureFile(50mm,50mm){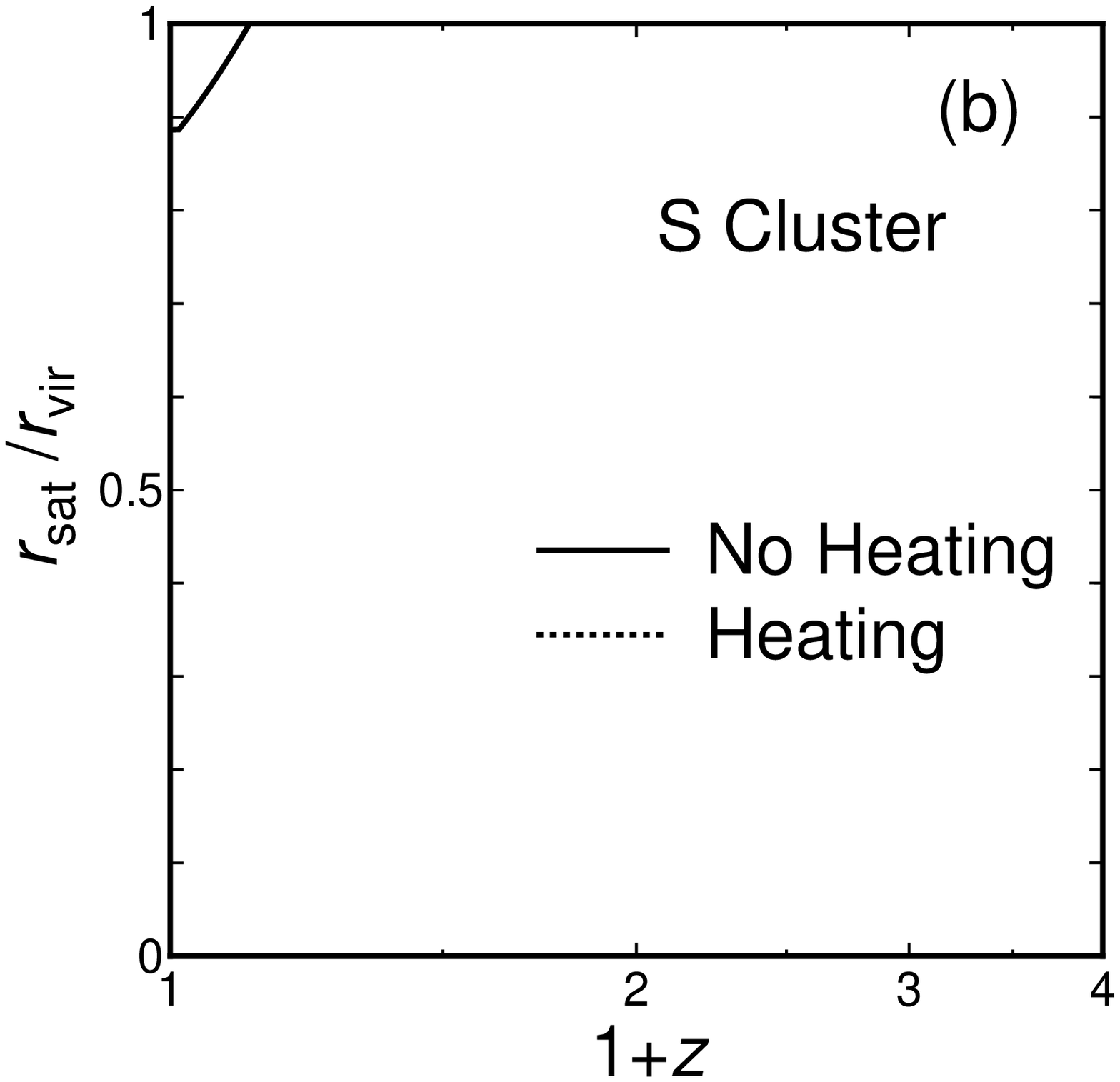}
  \end{center}
  \caption{Radii where thermal conduction is saturated, $r_{\rm sat}$,
 normalized by viral radii, $r_{\rm vir}$, for a bigger galaxy in a (a)
 main cluster, and (b) subcluster. The solid and dashed lines show the
 non-heated and the heated ICM models, respectively.  For $r<r_{\rm
 sat}$, thermal conduction is not saturated. For the model group,
 $r_{\rm sat}/r_{\rm vir}>1$ regardless of $z$.} \label{fig:rsat}
\end{figure}

\begin{figure}
  \begin{center}
    \FigureFile(50mm,50mm){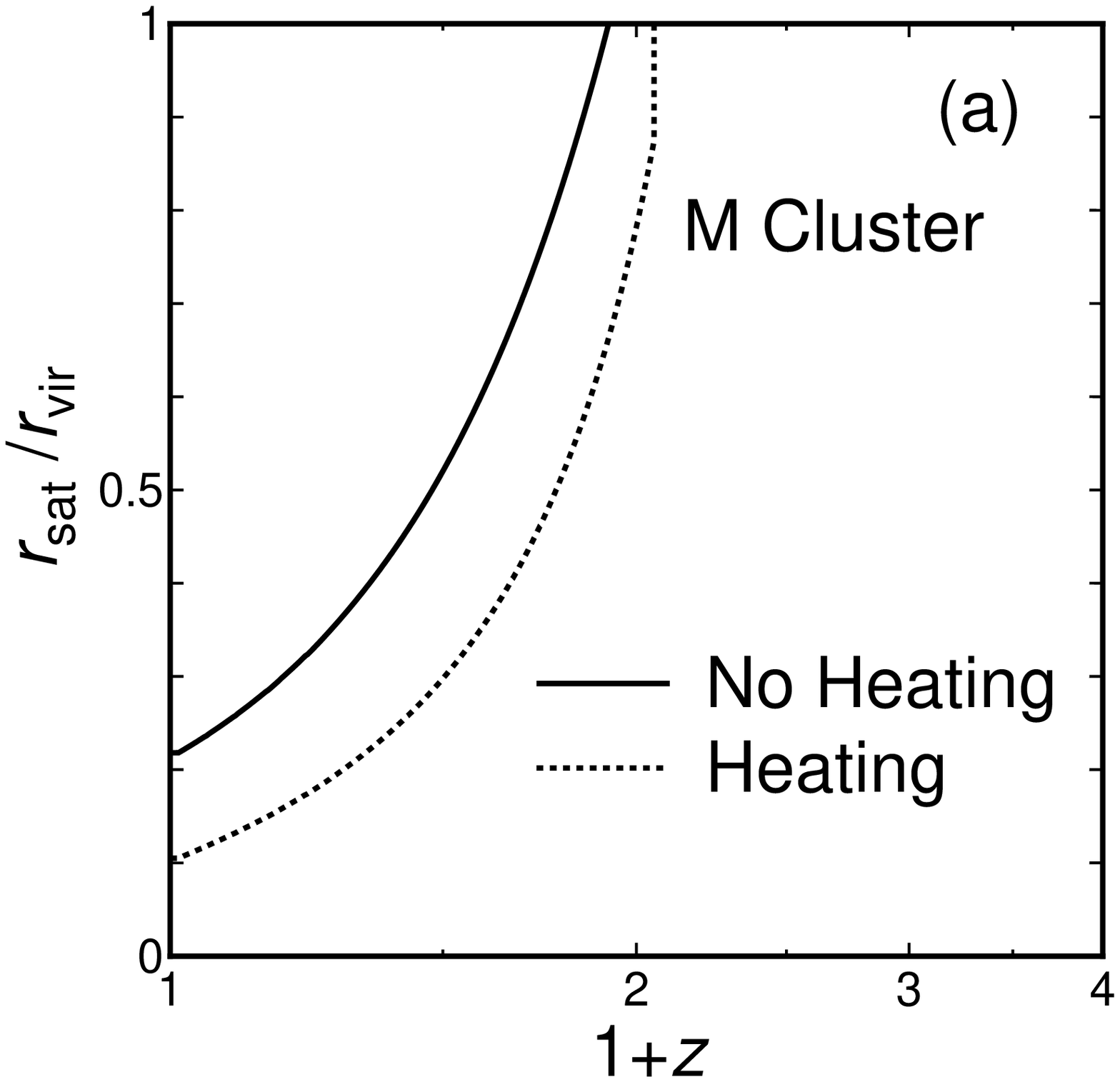}
    \FigureFile(50mm,50mm){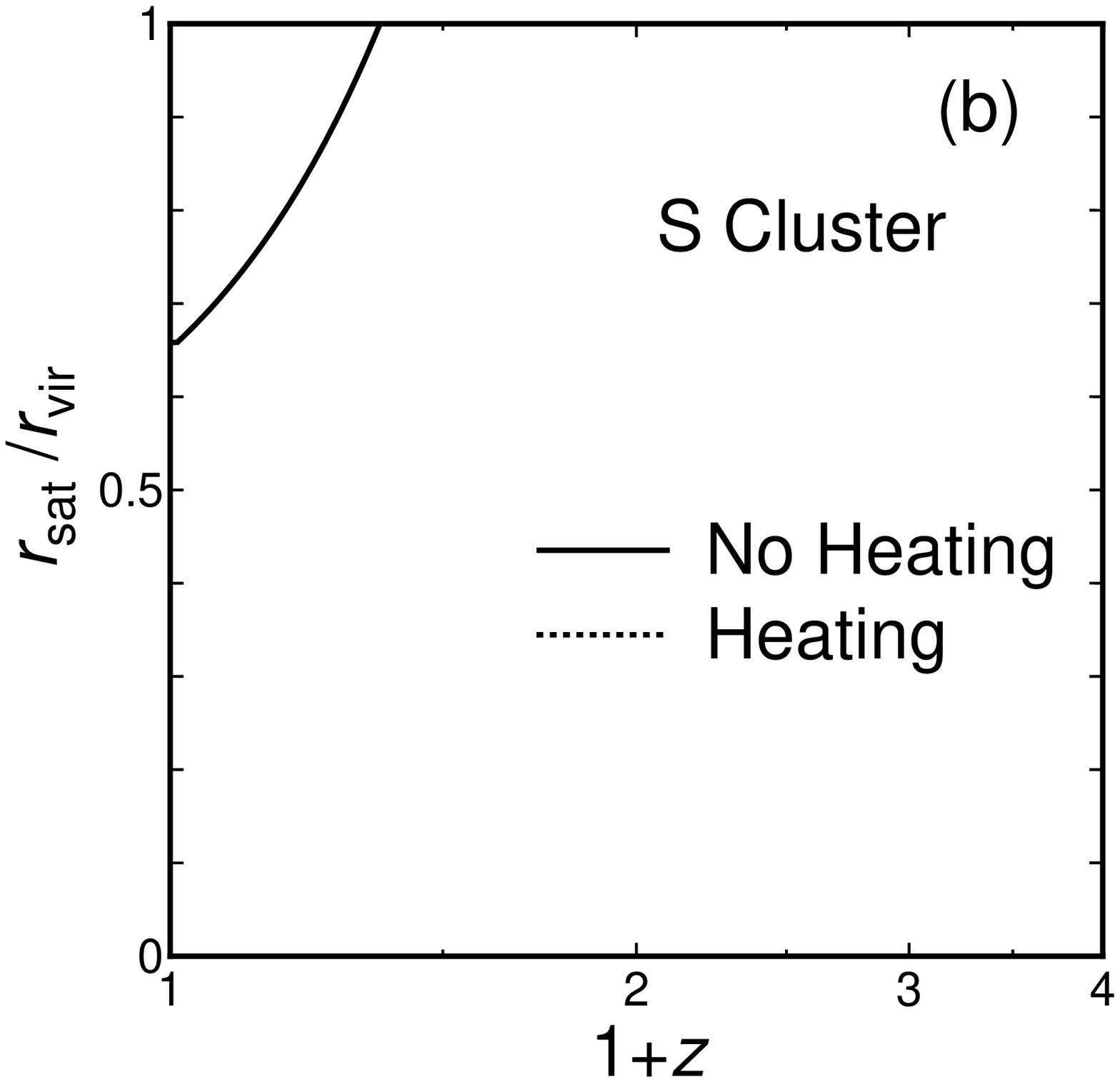}
  \end{center}
  \caption{Same as figure~\ref{fig:rsat},
 but for a smaller galaxy.} \label{fig:rsat_s}
\end{figure}

We investigate whether evaporation has an influence on galaxies being
{\it not} at the center of a cluster (or a subcluster or a
group). Figures~\ref{fig:rsat} and~\ref{fig:rsat_s} show $r_{\rm sat}$
normalized by $r_{\rm vir}$. For the main cluster, $r_{\rm sat}/r_{\rm
vir}<1$ at $z\lesssim 0.7$--$1$. Because of saturation, the energy flux
via thermal conduction is smaller than equation~(\ref{eq:Lcond}) at
$r>r_{\rm sat}$. The time scale of evaporation by the {\it saturated}
heat flux [equation~(\ref{eq:Lsat})] is $\sim 1$--$3$~Gyr around the
viral radius of the main cluster for $z\lesssim 0.7$--$1$.  Since the
time scale of evaporation is larger than $t_{\rm wst}$
(figure~\ref{fig:twst}), we expect that ram-pressure stripping becomes
effective before the evaporation affects the cold gas of a galaxy if the
galaxy has fallen directly from the outside of the cluster. (As
discussed in section~\ref{sec:disc}, this is inconsistent with the
observations of CNOC clusters.)  For the subcluster, $r_{\rm sat}/r_{\rm
vir}>1$ except for the non-heated ICM model at very low redshift
(figures~\ref{fig:rsat}b and~\ref{fig:rsat_s}b) and for the group,
$r_{\rm sat}/r_{\rm vir}$ is always larger than one. Therefore, we can
conclude that evaporation is effective in almost all regions of the
subcluster for $z\lesssim 0.5$ and in all regions of the group for
$z\sim 0$ if ram-pressure stripping is ignored
(figures~\ref{fig:tcond}b, \ref{fig:tcond}c, \ref{fig:tcond_s}b,
and~\ref{fig:tcond_s}c). As mentioned in subsection~\ref{sec:res_ram},
ram-pressure stripping may not occur in subclusters, if
non-gravitational heating alters the ICM distributions and the orbits of
galaxies are affected by tidal acceleration from the main cluster. In
this case, cold gas in a galaxy gradually disappears on a time scale of
$t_{\rm cond}\sim 2$--$7$~Gyr (for $z\lesssim 0.5$) without being affected
by ram-pressure stripping, while the galaxy circulates around the
subcluster center several times (figures~\ref{fig:tcond}b
and~\ref{fig:tcond_s}b). For the group, the time scale of evaporation
($t_{\rm cond}\sim 10$~Gyr at $z\sim 0$) is much larger than the
dynamical time scale of the group ($\sim 2$~Gyr at $z\sim 0$). Thus, we
may need to consider the evolution of the group before we conclude
evaporation of the galaxy in the group.

\subsection{Galaxy Mergers}

\begin{figure}
  \begin{center}
    \FigureFile(50mm,50mm){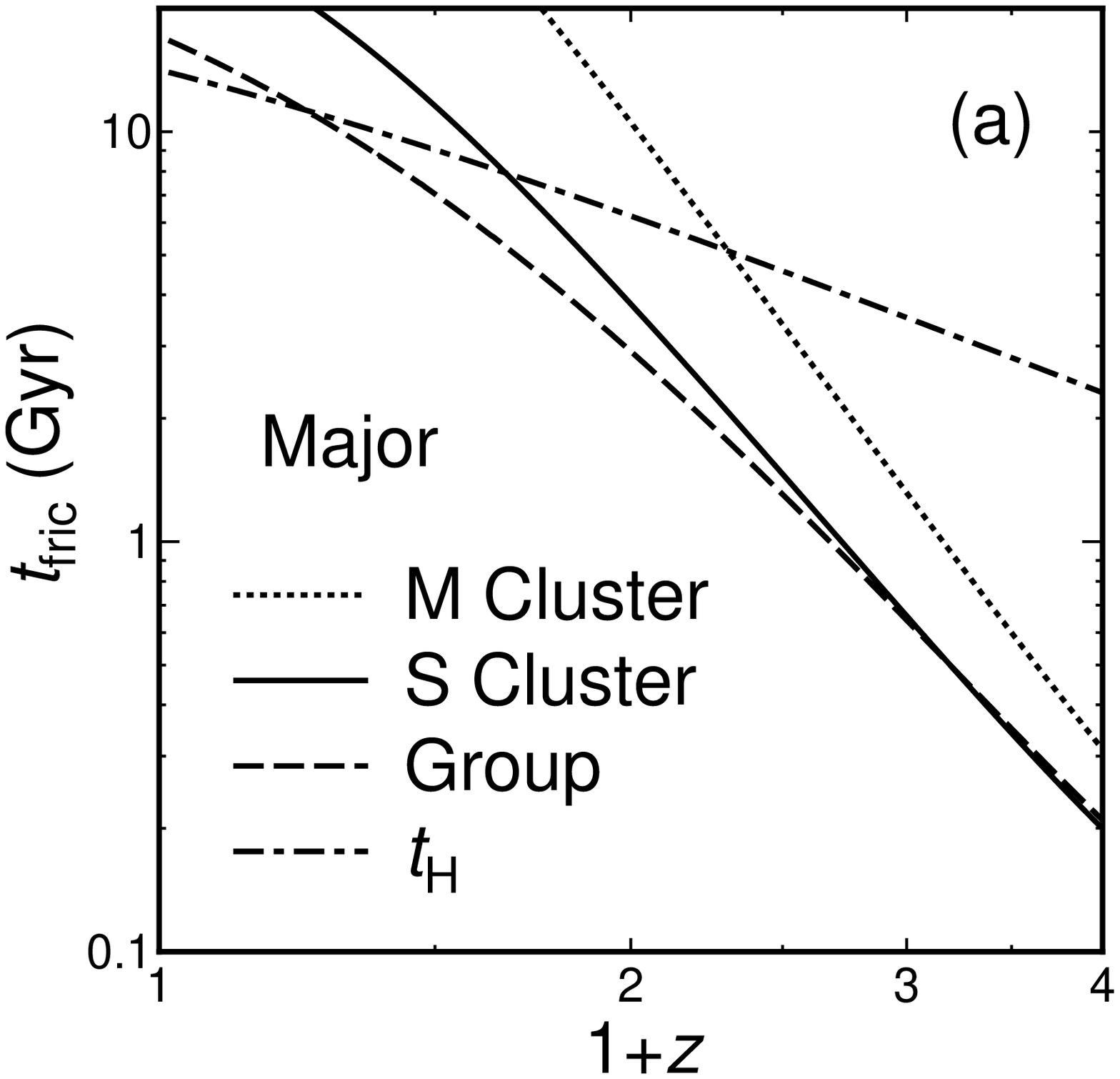}
    \FigureFile(50mm,50mm){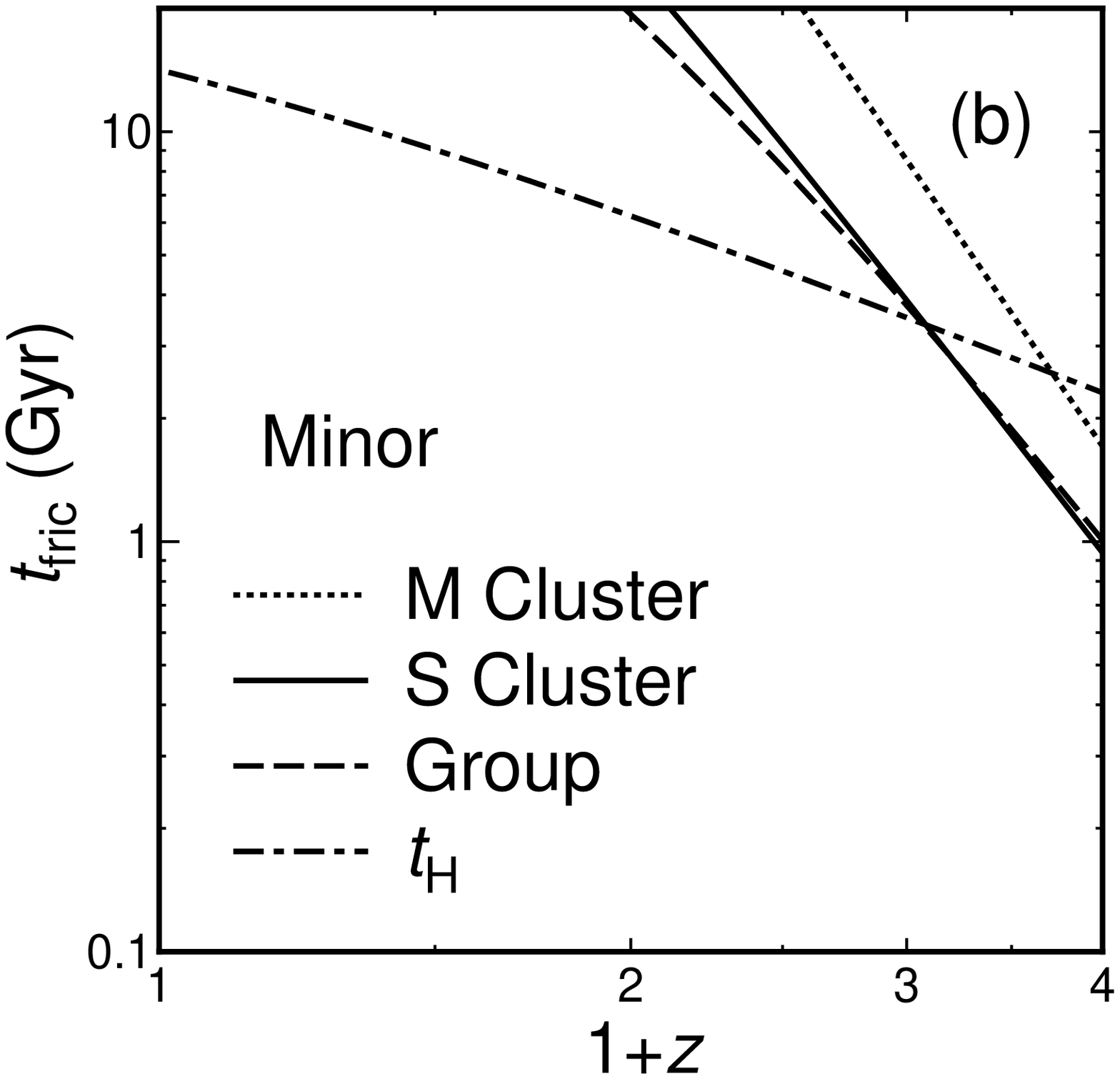}
  \end{center}
  \caption{Time scales of dynamical friction for a main cluster (dotted
line), a subcluster (solid line), and a group (dashed line). The mass of
a satellite galaxy is (a) $M_{\rm gal}=2\times 10^{11}\; M_\odot$, and
(b) $M_{\rm gal}=2\times 10^{10}\; M_\odot$. The mass of the central
galaxy should be larger than these values. The dot-dashed lines show the
Hubble time, $t_{\rm H}$.  For $t_{\rm fric}\lesssim t_{\rm H}$, galaxy
mergers are effective.} \label{fig:tfric}
\end{figure}

After a galaxy is caught by a massive dark halo, its orbit will be
affected by dynamical friction. The orbit gradually shrinks and the
galaxy merges with the central galaxy of the halo on a time scale of
$t_{\rm fric}$. As mentioned in subsection~\ref{sec:mod_merger}, we
consider mergers in which one of the merging galaxies is at the halo
center and that the mass is sufficiently large. We also assume that the
mass of another galaxy falling into the halo center is $M_{\rm
gal}=2\times 10^{11}\; M_\odot$ (figure~\ref{fig:tfric}a) and $2\times
10^{10}\; M_\odot$ (figure~\ref{fig:tfric}b). These masses include those
of the individual galactic halos surrounding the infalling galaxies. The
mass of the bigger galaxy ($2\times 10^{11}\; M_\odot$) is based on the
traditional Galaxy mass (e.g. \cite{all73}), although recent studies
show that the mass of the Galaxy is much larger \citep{sak03}. However,
most of the mass resides in the outer halo region, and it would tidally
be stripped in a cluster. Thus, we assumed the traditional value. In
figure~\ref{fig:tfric}, we present $t_{\rm fric}$ for our model cluster,
subcluster, and group. The Hubble time $t_{\rm H}$ is also
depicted. Galaxy mergers are effective when $t_{\rm fric}\lesssim t_{\rm
H}$. When $M_{\rm gal}=2\times 10^{11}\; M_\odot$ ($2\times 10^{10}\;
M_\odot$), we do not expect galaxy mergers at $z\lesssim 1.3$ (2.7) in
the main cluster and at $z\lesssim 0.6$ (2.1) in the subcluster.

Since the masses of most galaxies are smaller than $2\times 10^{11}\;
M_\odot$, the results indicate that galaxy mergers had already ceased in
the subclusters observed around massive clusters at $z\lesssim 0.6$. For
the main cluster, galaxy mergers have stopped at higher redshift, because
the velocities of galaxies are large in the
cluster. Figure~\ref{fig:tfric} suggests that galaxies at the centers of
massive clusters (cD galaxies) formed at $z\gtrsim 1$ if they formed
through galaxy mergers. For the group in figure~\ref{fig:tfric}a, galaxy
mergers had recently stopped at $z\sim 0.2$. Thus, the situation is
similar to that of the subclusters observed at $z\sim 0.5$.

\section{Discussion}
\label{sec:disc}

Star-formation histories of galaxies in clusters and subclusters may
tell us the relation between environmental effects and the truncation of
the star-formation activities of the galaxies. Unfortunately, there have
not been detailed observations about the star-formation histories of
galaxies in subclusters around the main clusters. Thus, we first discuss
the star-formation histories of galaxies in main clusters, and then
infer what has happened in the subclusters.

Sharp truncation of star formation ($\lesssim 10^9$~yr) would lead to an
abundant population of galaxies with strong Balmer lines, but no nebular
emission lines (K$+$A galaxies, but see \cite{mil01}). Large fractions
of K$+$A galaxies have been reported in some main clusters at
$0.4\lesssim z \lesssim 0.6$, in particular by the MORPHS collaboration
\citep{dre99,pog99}. On the other hand, these galaxies appear to be rare
in the CNOC sample of very luminous X-ray clusters studied by
\citet{bal99b}. Moreover, Kodama and Bower (2001) analyzed CNOC
clusters, and indicated that if star formation declines on a relatively
long time scale ($1-3$~Gyr) after accretion to the clusters, the galaxy
accretion history that they infer is consistent from cluster to cluster
and matches well the distribution of galaxies in the color-magnitude
diagram expected in simple theoretical models. On the other hand, models
in which star formation is abruptly truncated as galaxies are accreted
by the cluster have difficulty in reproducing the observed color
distribution. It is still unclear whether this apparent disagreement
between the MORPHS and the CNOC clusters is a result of the procedure
used to select the spectroscopic sample, the effects of dust
obscuration, or perhaps a genuine effect of the dependence of
star-formation rates on other cluster properties such as X-ray
luminosity or temperature \citep{bal99b}. In the following, we mostly
discuss those distant clusters observed at $z\sim 0.5$.

If the former observational results can be applied to most clusters,
ram-pressure stripping is the most promising candidate of the mechanism
that suppresses star formation, because ram-pressure stripping removes
cold gas in a galactic disk almost instantaneously \citep{aba99,qui00},
and as a result, the star-formation rate of the galaxy drops rapidly on
a time scale of $\sim 10^8$~yr \citep{fuj99a}. Moreover, ram-pressure
stripping naturally accounts for the observed trend that fractions of
blue galaxies are systematically lower for rich clusters than for poor
clusters at a given redshift (Paper~I; \cite{mar01,got03a}).

On the other hand, for the subclusters at $z\sim 0.5$, ram-pressure
stripping also seems to be important if the ICM has not been heated
non-gravitationally (figures~\ref{fig:rst}b
and~\ref{fig:rst_s}b). Although quantitative discussion may need the
knowledge of distribution of galaxy orbits, it is at least more
efficient than the groups with the same mass at $z\sim 0$. Ram-pressure
stripping in the subclusters could be observed as the existence of K$+$A
galaxies. If they are really observed, it would show that the
non-gravitational heating of the ICM has not occurred at that redshift
at least for the subclusters.

If the slow decline of star-formation rates of galaxies observed in some
main clusters is confirmed for most clusters, it would be difficult to
make a compromise with our models. Ram-pressure stripping is effective
regardless of non-gravitational heating in the main cluster at
$z\lesssim 1$ (figures~\ref{fig:rst}a and~\ref{fig:rst_s}a). Even if we
adopt a strangulation model, ram-pressure stripping starts before
strangulation finishes (subsection~\ref{sec:res_ram}), and it leads to a
sharp decline of the star-formation rates of galaxies. \citet{bek02}
demonstrated numerical simulations and claimed that warm gas in the halo
of a disk galaxy can gradually be stripped by ram-pressure from ICM
($\gtrsim$~Gyr). However, the orbit of a galaxy that they assumed is far
from a radial one in spite of the fact that galaxies falling into a
cluster for the first time take almost radial orbits.

One possible solution of the problem is that most galaxies had already
been affected by strangulation and/or evaporation to some degree in
subclusters before they entered the main clusters. If this is the case,
the total time in which the galaxies are under the influence of
strangulation or evaporation can be larger than $t_{\rm wst}$ shown in
figure~\ref{fig:twst}. Thus, strangulation and/or evaporation could be
completed before galaxies are affected by ram-pressure stripping,
especially when ram-pressure stripping is avoided in subclusters by
non-gravitational heating and tidal acceleration from the main cluster
(subsection~\ref{sec:res_ram}). In particular, the evaporation scenario
works even if all warm gas around a disk galaxy had fallen onto the
galactic disk at a relatively high redshift (see \cite{ben00}). We
emphasize that non-gravitational heating, tidal acceleration, or
something else had to affect the ICM distributions or the galaxy orbits
in subclusters in order to effectively avoid the intensive gas removal
through ram-pressure stripping. If cold gas in most galaxies had
completely stripped by ram-pressure stripping and their star-formation
activities had ceased until they entered main clusters, the expected
color distribution of galaxies in clusters may not be consistent with
the blue galaxy fractions of the CNOC clusters estimated by Kodama and
Bower (2001).

The above `pre-processing' scenario about the decline of star-formation
activities of galaxies is consistent with recent observations.
Morphological studies of distant cluster galaxies revealed the presence
of an unusual population of galaxies with a spiral morphology and the
lack of star-formation activity \citep{cou98,dre99,pog99}. These
galaxies are called `passive spiral galaxies'. In particular,
\citet{got03b} analyzed the Sloan Digital Sky Survey (SDSS) galaxy
catalogue and showed that passive spiral galaxies live at $\sim 1$--$10$
cluster-centric virial radius. These galaxies may be the galaxies
affected by strangulation and/or evaporation in subclusters around the
main clusters, because strangulation and evaporation affect the
star-formation activities of spiral galaxies, but not their
morphology. In fact, using Suprime-Cam on the Subaru Telescope,
\citet{kod01c} showed that the color change of galaxies occurs in
subclumps well outside the main cluster Abell~851. On the other hand, we
should note that the existence of passive spiral galaxies alone can be
explained by ram-pressure stripping in subclusters.

Galaxy mergers are thought to be responsible for the morphological
changes of galaxies, although they may also induce starburst. We assumed
that the mass of the central galaxy of a cluster or a subcluster is
larger than the mass of galaxies that are not at the cluster or
subcluster center (satellite galaxies) and is $\gtrsim 10^{11}\;
M_{\odot}$. Although the central galaxies of current massive clusters
are mostly massive elliptical galaxies, the central galaxies might be
massive spiral galaxies in the cluster progenitors when the masses of
the progenitors were small.

For the main cluster, mergers between massive galaxies or major mergers
are rare at $z\lesssim 1.3$ (figure~\ref{fig:tfric}a). This is because
as a cluster grows, the typical velocity of galaxies in the cluster
increases and $t_{\rm fric}$ decreases
[equation~(\ref{eq:fric})]. Mergers between central galaxies of clusters
and galaxies with small mass (minor mergers) had finished even earlier
(figure~\ref{fig:tfric}b). Thus, for the observed main clusters at
$z\lesssim 1$, mergers at the cluster centers should have
stopped. Compared with the main cluster, galaxy mergers could occur at a
lower redshift in the subclusters (figure~\ref{fig:tfric}). Major
mergers occur even at $z\sim 0.6$.

Major mergers would create elliptical galaxies \citep{too72}. In fact,
theoretical models including major mergers give an excellent explanation
for colors and distribution of cluster ellipticals
\citep{kau98,oka01,nag01,dia01}. Thus, our model for main clusters (and
many other previous studies) shows that the formation of elliptical
galaxies at $r\sim 0$ is at $z>1$. A main cluster might be divided into
several clusters at high redshift and elliptical galaxies observed at
$r\sim0$ at present might form in those main clusters. Since major
mergers could occur at lower redshift in subclusters
(figure~\ref{fig:tfric}a), they may be observed in the cluster
proximity, even at $z<1$. We note that using $N$-body simulations,
\citet{ghi98} showed that mergers between haloes in the cluster
proximity occur with a frequency of about $5$--$10$\% even since
$z=0.5$. However, it is not certain that this frequency corresponds to
the merger frequency of galaxies in those halos.

On the other hand, major mergers alone cannot account for the fraction
of galaxies with intermediate bulge-to-disk luminosity ratios such as S0
galaxies \citep{oka01,dia01}. Okamoto and Nagashima (2003) indicated
that minor mergers may create the galaxies with intermediate
bulge-to-disk luminosity ratios because minor mergers do not disrupt
galactic disks completely. However, they also indicated that the
fraction of the galaxies with intermediate bulge-to-disk luminosity
ratios does not evolve with a redshift contrary to observations
\citep{dre97}. This is consistent with our prediction that minor mergers
had finished at fairly high redshift for both the main cluster and the
subcluster ($z\gtrsim 2$; figure~\ref{fig:tfric}b). These studies
suggest that minor mergers in clusters are not the mechanism of the
observed morphological transformation to S0 galaxies at $z\lesssim 1$.

Ram-pressure stripping could darken a galactic disk and increase the
bulge-to-disk luminosity ratio of a disk galaxy \citep{fuj99a}. However,
the total luminosity of the galaxy may decrease too much to be observed
as a bright massive galaxy (\cite{oka03}, see also
\cite{bal02a}). Moreover, Kodama and Smail (2001) indicated that the
process responsible for the morphological transformation from a spiral
to a S0 galaxy takes a relatively long time ($\sim 1$--$3$~Gyr) after the
galaxy has entered the cluster environment. The time scale would be too
long for the morphological transformation by ram-pressure stripping
\citep{fuj99a}. Thus, the morphological transformation may be due to a
change of Q-parameter after strangulation \citep{bek02} or tidal
acceleration from a cluster potential on galaxies
\citep{byr90,val93,hen96,fuj98}. \citet{dal01} showed that early-type
spirals in cluster cores appear to be more perturbed than their
counterparts in the cluster peripheries by a factor of 2. This result
suggests that early-type spiral galaxies in clusters likely experienced
gravitationally induced disturbances as they pass near the cluster
cores, In a massive main cluster, cold gas of a disk galaxy may have to
be removed or consumed until the galaxy is affected by the tidal
acceleration in the central region of the cluster; otherwise the kinetic
pressure on the cold gas and the star-formation rate of the galaxy would
increase there, which is inconsistent with observations of massive
clusters \citep{fuj98}. A galaxy may have to pass the cluster (or
subcluster) center several times until the morphology is significantly
changed. Numerical simulations done by \citet{ghi98} showed that in a
cluster, galaxies follow almost radial orbits rather than circular ones
even after their first infalls to the cluster. It is to be noted that
\citet{gne03} indicated that the morphological transformation by tidal
force may occur even in the outer region of a hierarchically growing
cluster. Thus, some of observed S0 galaxies were formed outside the core
of a cluster. Recently, \citet{tre03} showed that the morphology-density
relation of galaxies holds even the outside of a cluster. This suggests
that the morphological change does not take place in field regions but
in subclusters.

Finally, we point out that some of blue disk galaxies observed in
clusters may be galaxies that restart star formation after they have
passed the central regions of the clusters. Although the cold gas of
these galaxies was once removed by ram-pressure stripping, the galaxies
could accumulate cold gas again from the gas ejected from their stars if
ram-pressure drops enough.  The condition that cold gas is accumulated
again is given by
\begin{eqnarray}
  \label{eq:ret}
   \rho_{\rm ICM}v_{\rm rel}^2 
  &< & \frac{16}{v_{\rm rel}}  
 \frac{S}{\pi r_{\rm gal}^2}v_{\rm rot}^2 \nonumber\\
  &= & 2.0\times 10^{-11} {\rm dyn\; cm^{-2}}
    \left(\frac{v_{\rm vel}}{500\rm\; km\; s^{-1}}\right)^{-1} 
       \nonumber\\
  &  & \times
    \left(\frac{S}{6\; M_\odot\rm \;yr^{-1}}\right)
    \left(\frac{r_{\rm gal}}{10\rm\; kpc}\right)^{-2}
    \left(\frac{v_{\rm rot}}{220\rm\; km\; s^{-1}}\right)^{2}
\;,
\end{eqnarray}
where $S$ is the gas ejection rate from stars
\citep{tak84,fuj99b}. Comparing equation~(\ref{eq:strip}) with
equation~(\ref{eq:ret}), one finds that the threshold values are very
similar for a massive disk galaxy we study. Thus, the gas accumulation
and star formation would resume at $r\sim r_{\rm st}$ after the galaxy
passed the central region of a cluster. A detailed discussion of the
resumed star formation is beyond the scope of this paper, because the
orbit of a galaxy that has passed the center of a cluster is gradually
affected by gravity from other galaxies and changing potential well of
the host cluster. Numerical simulations are needed to follow the orbits
of galaxies for a long time. We note that using a numerical simulation,
\citet{fuj99b} showed that the star formation of galaxies is suppressed
during a cluster merger, because the ram-pressure on galaxies increases
and ram-pressure stripping becomes effective. However, they also showed
that after a smaller cluster passed the center of a larger cluster, star
formation resumes because of the decrease of ram-pressure and the
accumulation of cold gas in the galaxies. Moreover, from $N$-body
numerical simulations of cluster formation, \citet{bal00} indicated that
a significant fraction of galaxies beyond the virial radius of a cluster
may have been within the main body of the cluster in the past. These
galaxies would resume the star-formation activities.

\section{Summary and Conclusions}
\label{sec:sum}

Using analytical models based on a hierarchical clustering scenario, we
have investigated the evolution of massive disk galaxies in clusters. We
have considered the effects of cosmological evolution of clusters on
galaxy evolution. In particular, we have explored the galaxy evolution
in subclusters located around the main cluster. The main conclusions
from this work may be summarized as follows:

(i) As long as the spatial distribution of intracluster medium (ICM)
follows that of dark matter in a cluster or a subcluster, galaxies are
influenced by ram-pressure stripping of the cold disk gas. For a given
massive cluster at $z=0$, ram-pressure stripping in the cluster
progenitors is more effective at higher redshift although the masses of
the cluster progenitors decrease with redshift. This is because the
typical mass density of the progenitors increases with the
redshift. Even in subclusters, ram-pressure stripping is effective.

(ii) If field galaxies directly fall into a main cluster, the exhaustion
of the cold disk gas via star formation, after stripping of the warm
diffuse gas in their galactic halos (`strangulation'), would not be
completed before ram-pressure stripping of the cold disk gas becomes
effective. This limits the time scale in which star-formation rates of
the galaxies decline. The maximum time scale is $\lesssim$~Gyr, which is
inconsistent with some observations.

(iii) The conflict between the theoretical prediction and the
observations may be avoided if the star-formation rates of galaxies had
already dropped in subclusters before the subclusters plunged into the
main cluster (`pre-processing'). This is consistent with recent
observations indicating that the star-formation rates of galaxies are
decreased at radii well larger than the virial radii of main clusters.

(iv) The star-formation rates of galaxies in subclusters might have
gradually decreased via strangulation and/or evaporation of the cold gas
by the surrounding hot ICM, if non-gravitational heating of ICM
(`preheating') had changed the ICM distributions of the subclusters, and
tidal force from the main cluster prevents the galaxies from being
affected by ram-pressure stripping. These may be the mechanism of the
pre-processing. In particular, the evaporation scenario is free from the
problem that warm gas in galactic halos has not been detected.

(v) It is not obvious whether ram-pressure stripping takes place in
subclusters (or main clusters at high redshift) or not. Even if future
observations show that ram-pressure stripping does not occur in
subclusters (or main clusters at high redshift), their small masses
cannot be the only reason. It requires additional reasons, such as the
preheating and/or the tidal acceleration.

(vi) Mergers between a galaxy at the center of a cluster progenitor and
that not at the center had mostly finished at $z\gtrsim 1$--$2$ because
the velocities of galaxies increase as the host clusters grow. Thus,
galaxy mergers do not appear to be the main cause of the observed
morphological transformation from spiral galaxies to S0 galaxies at
$z\lesssim 1$.

(vii) The star-formation activity of a galaxy may resume after the
galaxy has passed the central region of a cluster and the ram-pressure
from the ICM has dropped. This may affect the fraction of blue galaxies
in clusters.

\vspace{10mm}

I am grateful to M. Nagashima, T. Okamoto, T. Kodama, N. Yoshida, and
T. Goto for useful discussion.  I was supported in part by
Grants-in-Aid from the Ministry of Education, Culture, Sports, Science
and Technology (14740175).

\end{document}